\documentclass[aps,prd,showpacs,superscriptaddress,preprint,raggedfooter]{revtex4-1}   
\usepackage{graphicx}  
\usepackage[utf8]{inputenc}

\usepackage{dcolumn}   
\usepackage{array}     
\newcolumntype{K}[1]{>{\centering\arraybackslash}p{#1}} 
\usepackage{bm}        
\usepackage{amssymb}   
\usepackage{amsmath}   
\usepackage{mathrsfs}  
\usepackage{subfigure}
\usepackage{hyperref}
\usepackage{color}

\usepackage[english]{babel}

\newcommand{\lsim}{\stackrel{\scriptstyle <}{\phantom{}_{\sim}}}
\newcommand{\gsim}{\stackrel{\scriptstyle >}{\phantom{}_{\sim}}}

\begin{document}

\title{Multi-kink scattering in the double sine-Gordon model}

\author{Vakhid A. Gani}
\email{vagani@mephi.ru}
\affiliation{Department of Mathematics, National Research Nuclear University MEPhI (Moscow Engineering Physics Institute), 115409 Moscow, Russia}
\affiliation{Theory Department, National Research Center Kurchatov Institute, Institute for Theoretical and Experimental Physics, 117218 Moscow, Russia}
\author{Aliakbar Moradi Marjaneh}
\email{Corresponding author, moradimarjaneh@gmail.com}
\affiliation{Young Researchers and Elite Club, Quchan Branch, Islamic Azad university, Quchan, Iran}
\author{Danial Saadatmand}
\email{saadatmand.d@gmail.com}
\affiliation{Department of Physics, University of Sistan and Baluchestan, Zahedan, Iran}

\begin{abstract}
We study collisions of two, three, and four kinks of the double sine-Gordon model. The initial conditions are taken in a special form in order to provide collision of all kinks in one point. We obtain dependences of the maximal energy densities on the model parameter. We also analyze the final states observed in these collisions.
\end{abstract}



\maketitle

\section{Introduction}\label{sec:introduction}

Interactions of kinks --- solutions of the type of solitary waves of non-integrable field-theoretical models in (1+1)-dimensional space-time --- are of growing interest \cite{Rajaraman.book.1982,Vilenkin.book.2000,Manton.book.2004,Vachaspati.book.2006}. Great amount of important results has been obtained recently for the kink-(anti)kink scattering \cite{Gani.PRD.2014,Gani.JHEP.2015,Gani.JPCS.2016.phi8,Bazeia.EPJC.2018.sinh,Bazeia.JPCS.2017.sinh,Gani.EPJC.2018.dsg,Belendryasova.JPCS.2019.dsg,Gomes.JHEP.2018,Bazeia.2018.hybrid,Dorey.PLB.2018.quasinormal,Campos.2019.quasinormal,Zhong.2019.inner}. In many cases rather non-trivial picture of interaction has been observed. In particular, it was found that the collision of kink and antikink crucially depends on the initial velocity $v_\mathrm{in}^{}$ (in the numerical experiments kink and antikink are initially placed at large distance from each other and are moving towards each other with the initial velocities $v_\mathrm{in}^{}$ in the laboratory frame of reference). There is a critical value of the initial velocity $v_\mathrm{cr}^{}$, which separates two different regimes of the collision. At $v_\mathrm{in}^{}>v_\mathrm{cr}^{}$ inelastic reflection of kinks (or passing through each other, if the model allows such process) is observed. At $v_\mathrm{in}^{}<v_\mathrm{cr}^{}$ after the collision the kink and antikink form a long-living bound state --- a bion --- which decays slowly radiating energy in the form of small-amplitude waves. Besides, in some models at $v_\mathrm{in}^{}<v_\mathrm{cr}^{}$ an interesting phenomenon was found --- so-called escape windows. Escape windows are intervals of the initial velocity, at which kinks do not form a bion, but escape from each other after two or more collisions. The cause of this phenomenon is the resonance energy exchange between translational and vibrational modes of the kink (antikink), see, {\it e.g.}, Refs.~\cite{Campbell.PhysD.1986,Kudryavtsev.UFN.1997.eng,Kudryavtsev.UFN.1997.rus}. In some models the kinetic energy of the kinks can be stored in vibrational modes of the system ``kink+antikink'' \cite{Gani.PRE.1999,Dorey.PRL.2011,Belendryasova.CNSNS.2019.phi8,Belendryasova.JPCS.2017.phi8}.

Another very interesting subject that should be mentioned here is interactions of kinks having power-law tails, see, {\it e.g.}, Refs.~\cite{Belendryasova.CNSNS.2019.phi8,Belendryasova.JPCS.2017.phi8,Gomes.PRD.2012,Radomskiy.JPCS.2017.tails,Christov.PRD.2019.long-range,Christov.2018.long-tails,Manton.2018.long-range,Manton.JPA.2018.long-range,Bazeia.JPC.2018.polynomial-tails}. Such kinks arise in various models with both polynomial and non-polynomial potentials. The power-law asymptotics (``fat tails'') of kinks result in their long-range interaction, {\it i.e.}\ the kinks feel each other at very large distances. As a result, the traditionally used initial conditions in the form of superposition of kink and antikink seem to be not valid for simulation of the kink-antikink scattering \cite{Christov.PRD.2019.long-range}.

The interactions of kinks are studied both numerically and analytically. On the one hand, the scattering processes can be simulated directly by using the numerical methods of solving equations of motion -- partial differential equations of second order. On the other hand, there are analytical approximate methods which allow to estimate forces between kink and (anti)kink. These are (i) the collective coordinate approximation \cite{Gani.PRD.2014,Weigel.PRD.2016.cc,Demirkaya.JHEP.2107.cc} and (ii) Manton's method \cite[Ch.~5]{Manton.book.2004}, \cite{Perring.NuclPhys.1962,Rajaraman.PRD.1977,Manton.NPB.1979,Kevrekidis.PRE.2004}.

It is worth to mention that kink-like solutions can also be obtained in more complicated models with two or more scalar fields \cite{GaKu.SuSy.2001.eng,GaKu.SuSy.2001.rus,Lensky.JETP.2001.eng,Lensky.JETP.2001.rus,Alonso-Izquierdo.PRD.2002,Kurochkin.CMMP.2004.eng,Kurochkin.CMMP.2004.rus,Alonso-Izquierdo.AHEP.2013.kink-like,Alonso-Izquierdo.JHEP.2014.moduli,Katsura.PRD.2014,Alonso-Izquierdo.Physica_D.2018.dynamics,Alonso-Izquierdo.PRD.2018.collisions,Alonso-Izquierdo.2018.MSTB,Alonso-Izquierdo.2019.scattering,Morris.AP.2018.mixing,Alonso-Izquierdo.2019.non-topological,Mohammadi.CNSNS.2019.two-field-sg}. In particular, in \cite{GaKu.SuSy.2001.eng,GaKu.SuSy.2001.rus,Alonso-Izquierdo.PRD.2002,Alonso-Izquierdo.Physica_D.2018.dynamics,Alonso-Izquierdo.PRD.2018.collisions,Alonso-Izquierdo.2018.MSTB,Alonso-Izquierdo.2019.scattering,Morris.AP.2018.mixing,Alonso-Izquierdo.2019.non-topological} the models with two real scalar fields with polynomial potentials have been studied. In \cite{GaLiRa,GaLiRaconf,blyankinshtein,GaKsKu01.eng,GaKsKu01.rus,GaKsKu02.eng,GaKsKu02.rus} configurations of the type of domain wall with additional fields localized on it were investigated. Many important results have also been obtained for non-topological field configurations --- lumps, Q-balls, etc.\ \cite{Brihaye.PRD.2015,Loginov.YadFiz.2011.eng,Loginov.YadFiz.2011.rus,Bazeia.EPJC.2017,Bazeia.fermion.2017,Schweitzer.PRD.2012.1,Schweitzer.PRD.2012.2,Schweitzer.NPA.2016,Nugaev.PRD.2013,Bazeia.EPJC.2016.Q-balls,Bazeia.PLB.2016.Q-balls,Bazeia.PLB.2017.Q-balls,Levkov.JHEP.2017.Q-balls,Loiko.PRD.2018.Q-balls,Kovtun.PRD.2018.Q-balls,Loginov.PLB.2018.Q-balls,Loginov.arXiv.2019.Q-balls}.

Kinks, domain walls and other topological defects arise in a great amount of models, hence they are very important for various physical applications from condensed matter to high energy physics and cosmology \cite{poltis1,valle2016,Matsui.JCAP.2016,Wachter.PRL.2017,Ye.EPJC.2017,Ringeval.JCAP.2017,GaKiRu,GaKiRu.conf}, see also \cite{Vilenkin.book.2000}. Many physical phenomena could be effectively described by one-dimensional topological structures. For example, a three-dimensional domain wall in the direction perpendicular to it can be viewed as a one-dimensional topological field configuration --- kink connecting two different vacua of the model. At the same time, the case of $(1+1)$-dimensional space-time structure can be investigated more easily compared to $2+1$ or $3+1$ dimensions.

The processes of collision of several kinks in one point (with extracting maximal energy densities) have been actively investigated over the last few years \cite{saadatmand.prd.2015,Moradi.JHEP.2017.multikink,Moradi.EPJB.2017.multikink,Moradi.CNSNS.2017.multikink,saad.WaveMotion.2018}. Such processes are of interest for numerous applications. Being topological defects, the kinks bear significant amount of energy. As a result, simultaneous collision of several kinks (antikinks) in one point can lead to formation  of spatial domains with high energy density. The authors of Ref.~\cite{saad.WaveMotion.2018} have shown that with knowing maximal energy densities in collision of $N$ kinks one can predict how many kink-antikink pairs can be
produced in particle collisions.

This study deals with the collisions of several kinks of the double sine-Gordon (DSG) model in one point. This model has been actively studied due to its numerous applications, see, {\it e.g.}, \cite{Gani.EPJC.2018.dsg,Gani.PRE.1999,Campbell.PhysD.1986.dsg,Malomed.PLA.1989.dsg,Nazifkar.BJP.2010.dsg} and references therein. Note that the subject of multi-kink collisions is very vast. The number of parameters that could be changed is very large. These are, in particular, the model parameter $R$, the initial velocities and the initial positions of kinks. In this paper we study collisions of two three and four kinks in one point (in a small spatial region). We focus on the dependences of maximal energy densities on the model parameter $R$. In collisions of four kinks (two kinks and two antikinks, to be more accurate) we investigate how the final configurations depend on $R$. In our numerical simulations we use fixed initial positions and initial velocities of kinks.

Our paper is organized as follows. In Sect.~\ref{sec:model} we recall the main features of the DSG model, write down its kinks, and discuss their main properties. Section \ref{sec:scattering} presents the results of the numerical simulations of multi-kink collisions. We conclude with summarizing in Sect.~\ref{sec:conclusion}, where we also formulate some possible directions for further research.


\section{Double sine-Gordon model}\label{sec:model}

Within the $(1+1)$-dimensional double sine-Gordon model the dynamics of the real scalar field $\phi(x,t)$ is described by the Lagrangian density
\begin{equation}\label{eq:Largangian}
\mathcal{L} = \frac{1}{2}\left(\frac{\partial\phi}{\partial t}\right)^2 - \frac{1}{2}\left(\frac{\partial\phi}{\partial x}\right)^2 - V(\phi),
\end{equation}
where the potential $V(\phi)$ is non-negative function of the field. From the Lagrangian \eqref{eq:Largangian} one can obtain the equation of motion for the field $\phi(x,t)$:
\begin{equation}\label{eq:eqmo}
\frac{\partial^2\phi}{\partial t^2} - \frac{\partial^2\phi}{\partial x^2} + \frac{dV}{d\phi} = 0.
\end{equation}
Below we simulate multi-kink scattering by solving this equation numerically.

\begin{figure*}[t!]
\begin{center}
  \centering
  \subfigure[]{\includegraphics[width=0.45
 \textwidth]{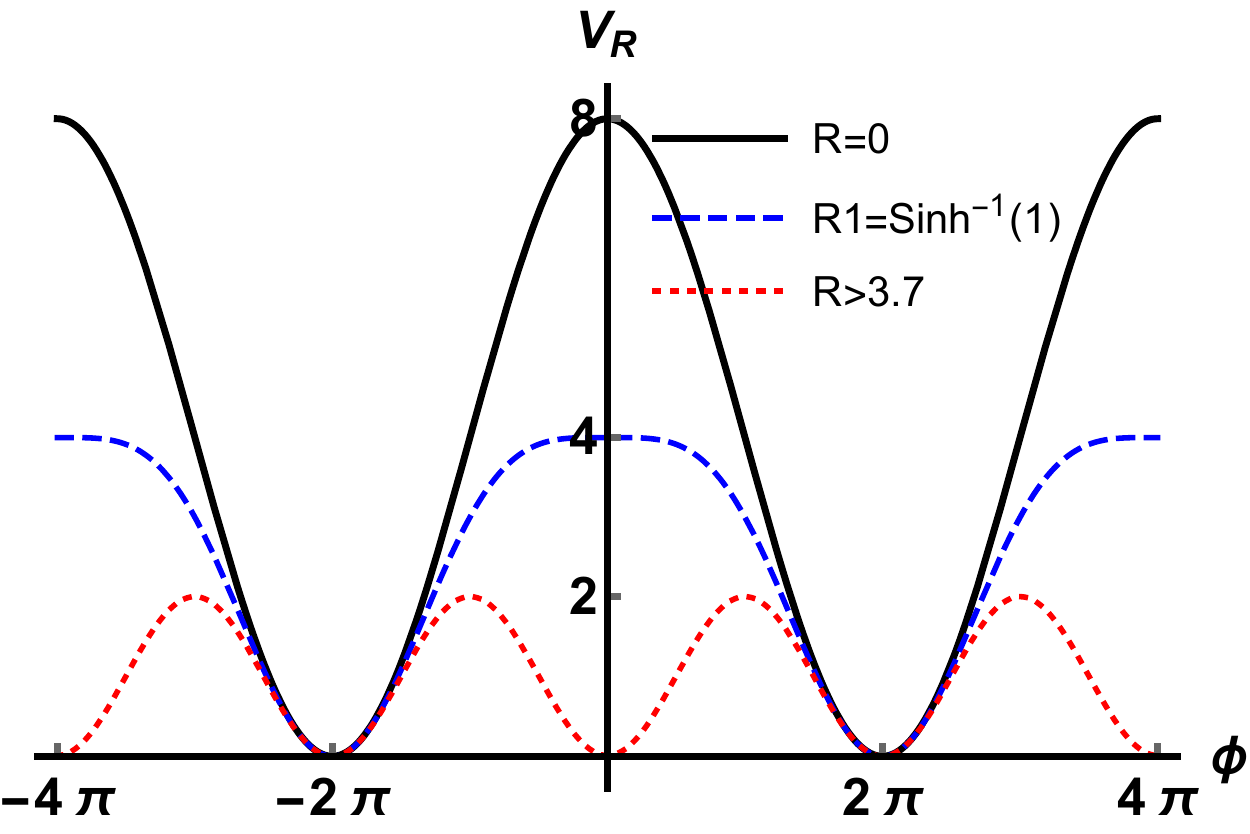}\label{fig:PotentialAsPhi}}
  \subfigure[
  ]
{\includegraphics[width=0.45\textwidth]{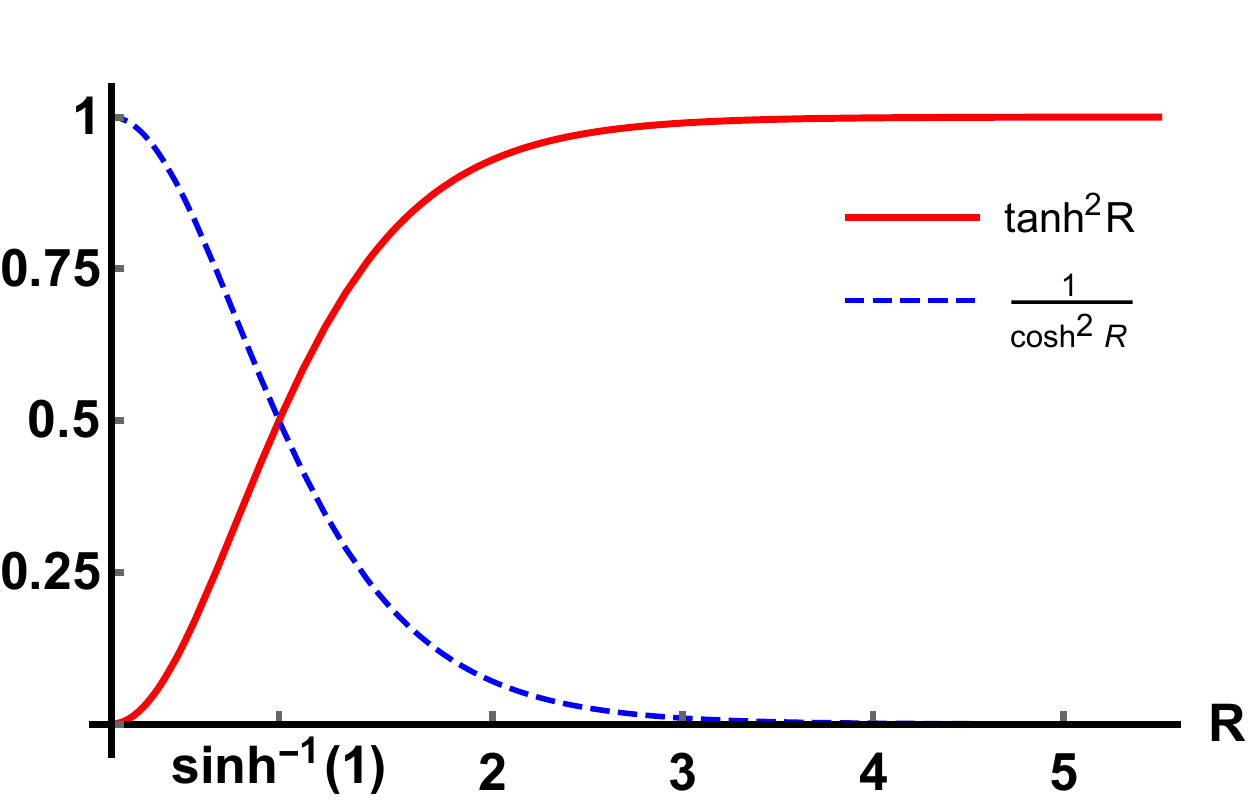}\label{fig:PotentialCoefficients}}
  \\
  \caption{(a) Potential of the DSG model; (b) $R$-dependence of the coefficients in the potential, which are responsible for the mixing of two sine-Gordon potentials.}
  \label{fig:Potential1}
\end{center}
\end{figure*}

In literature there are various modifications and parameterizations of the potential, which all can be called ``double sine-Gordon potential''. We will use the so-called $R$-parameterization, see, {\it e.g.}, Refs.~\cite{Gani.EPJC.2018.dsg,Gani.PRE.1999,Campbell.PhysD.1986.dsg}. In this parameterization the DSG potential is:
\begin{equation}\label{eq:potential_R}
V(\phi) = \tanh^2R\:(1-\cos\phi) + \frac{4}{\cosh^2R}\left(1+\cos\frac{\phi}{2}\right).
\end{equation}
In Fig.~\ref{fig:Potential1} we show (a) this potential for some values of the parameter and (b) the $R$-dependence of the coefficients in front of $(1-\cos\phi)$ and $\displaystyle\left(1+\cos\frac{\phi}{2}\right)$, which are responsible for the ``mixing'' of $\phi$-- and $\displaystyle\frac{\phi}{2}$--sine-Gordon potentials.
Depending on $R$, the potential \eqref{eq:potential_R} looks different. In particular,
\begin{equation}\label{eq:potential_R_limits}
V(\phi) =
\begin{cases}
4\left(1+\cos\displaystyle\frac{\phi}{2}\right)\quad \mbox{for}\quad R=0,\\
4-\left(1-\cos\displaystyle\frac{\phi}{2}\right)^2\quad \mbox{for}\quad R=\mbox{arsinh}\:1,\\
1 - \cos\phi\quad \mbox{for}\quad R\to+\infty.
\end{cases}
\end{equation}

\begin{figure}[t!]
\centering
\includegraphics[width=0.45\textwidth]{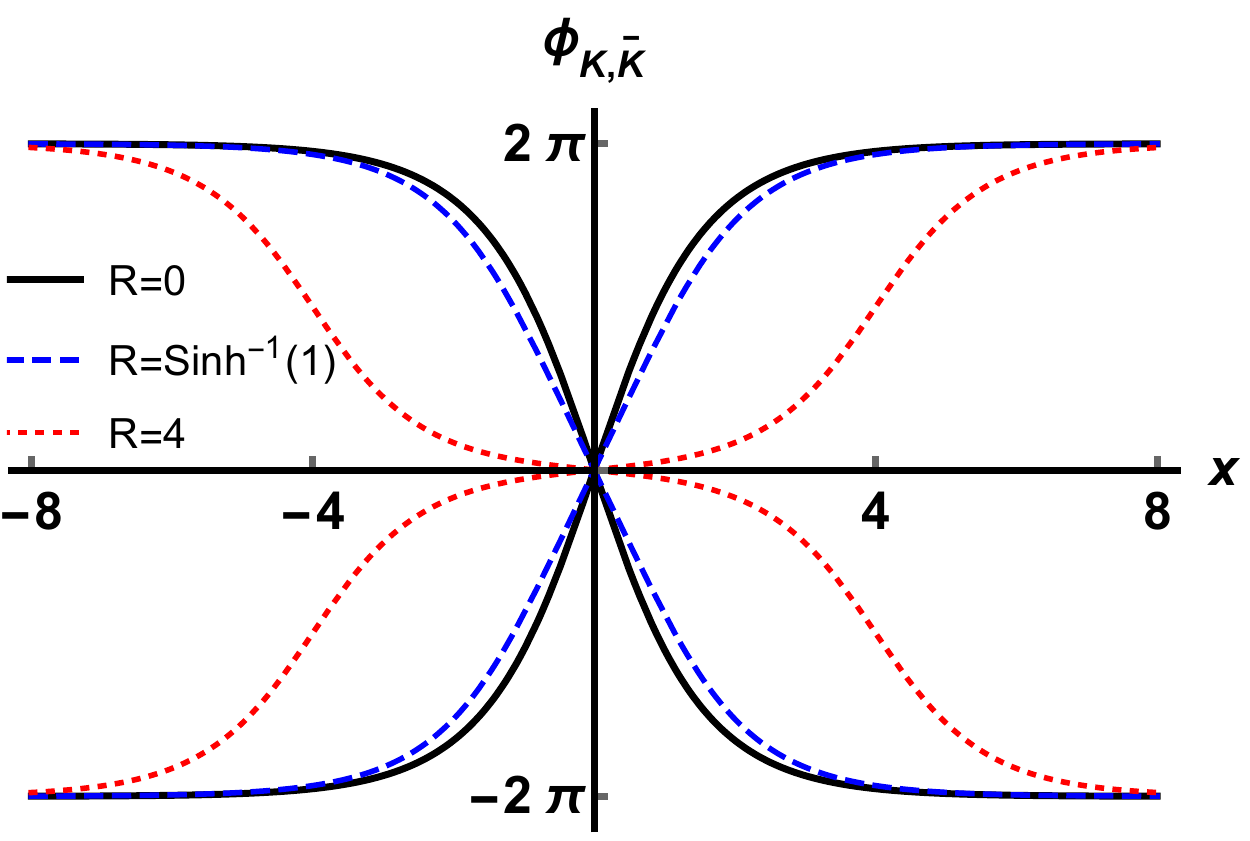}
\caption{Kinks (monotonously increasing functions $\phi_\mathrm{k}^{}(x)$) and antikinks (monotonously decreasing functions $\phi_\mathrm{\bar k}^{}(x)$) of the DSG model for three different values of the parameter $R$.} 
\label{fig:kinkantikinkSolutions}
\end{figure}

The double sine-Gordon model has static topological solution --- kink (antikink):
\begin{equation}\label{eq:DSG_kinks_1}
\phi_{\mathrm{k}(\mathrm{\bar k})}(x) = 4\pi n \pm 4\arctan\frac{\sinh x}{\cosh R}.
\end{equation}
Notice that for better understanding of many processes in kinks collisions the following fact could be important: the DSG kink (antikink) \eqref{eq:DSG_kinks_1} can be interpreted as a superposition of two sine-Gordon solitons:
\begin{equation}\label{eq:DSG_kinks_2}
\phi_{\mathrm{k}(\mathrm{\bar k})}^{}(x) = 4\pi n \pm \left[\phi_\mathrm{SGK}^{}(x+R)-\phi_\mathrm{SGK}^{}(R-x)\right],
\end{equation}
where
\begin{equation}
\phi_\mathrm{SGK}^{}(x)=4\arctan\exp (x)
\end{equation}
is the sine-Gordon kink (soliton). Thus the DSG kink can be viewed as a superposition of two sine-Gordon kinks, which are separated by the distance $2R$ (if the DSG kink is centered at $x=0$ then the two sine-Gordon solitons are centered at $x=\pm R$), see Fig.~\ref{fig:kinkantikinkSolutions}.
Due to the Lorentz invariance, the kink (antikink) moving along the $x$-axis with the velocity $v$ can be obtained by the Lorentz boost:
\begin{equation}
\phi_{\mathrm{k}(\mathrm{\bar k})}^{}(x,t) = \phi_{\mathrm{k}(\mathrm{\bar k})}^{}(\gamma(x-vt)),
\end{equation}
where $\gamma = 1/\sqrt{1-v^2}$ is the Lorentz factor.

The energy functional which corresponds to the Lagrangian \eqref{eq:Largangian} reads
\begin{equation}\label{eq:energy}
E[\phi] = \int\limits_{-\infty}^{\infty}\left[\frac{1}{2}\left(\frac{\partial\phi}{\partial t}\right)^2 + \frac{1}{2}\left(\frac{\partial\phi}{\partial x}\right)^2 + V(\phi)\right]dx.
\end{equation}
The total energy \eqref{eq:energy} can be split into three parts: (i) the kinetic energy $K$, (ii) the gradient energy $U$, and (iii) the potential energy $P$:
\begin{equation}
E = K + U + P.
\end{equation}
According to this, the integrand in \eqref{eq:energy}, {\it i.e.}\ the total energy density $\varepsilon(x,t)$, can be written as
\begin{equation}
\varepsilon(x,t) = k(x,t) + u(x,t) + p(x,t),
\end{equation}
where
\begin{equation}
k(x,t) = \frac{1}{2}\left(\frac{\partial\phi}{\partial t}\right)^2, \quad
u(x,t) = \frac{1}{2}\left(\frac{\partial\phi}{\partial x}\right)^2, \quad
p(x,t) = V(\phi)
\end{equation}
are (i) the kinetic energy density, (ii) the gradient energy density, and (iii) the potential energy density, respectively.

In the case of one moving kink $\phi_\mathrm{k}^{}(x,t)$ we have:
\begin{equation}
p(x,t) = \frac{8\cosh^2R\cdot\cosh^2 [\gamma (x-vt)]}{\left(\cosh^2R+\sinh^2[\gamma(x-vt)]\right)^2} = (1-v^2)\,u(x,t) = \frac{1-v^2}{v^2}\,k(x,t) = \frac{1-v^2}{2}\,\varepsilon(x,t).
\end{equation}
%
Integrating the total energy density with respect to $x$, we obtain the total energy of the moving kink:
\begin{equation}
E_\mathrm{K} = \int_{-\infty}^{+\infty}\varepsilon(x,t)\:dx = \frac{M_\mathrm{K}}{\sqrt{1-v^2}},
\end{equation}
where
\begin{equation}\label{eq:DSG_kink_energy}
M_\mathrm{K} = 16\left(1+\frac{2R}{\sinh 2R}\right)
\end{equation}
is the mass of kink, {\it i.e.}\ the energy of the static DSG kink.

In the next section we present the results of the numerical simulation of collisions of two, three and four DSG kinks in the same point. We find the maximal (over the spatial coordinate $x$ and time $t$) values of the energy densities mentioned above: kinetic, gradient, potential and total. Note that for one moving kink the maximal values can be obtained analytically. Depending on the parameter $R$, they are the following:
\begin{equation}
p_\mathrm{max}^{(1)} = (1-v^2)\,u_\mathrm{max}^{(1)} = \frac{1-v^2}{v^2}\,k_\mathrm{max}^{(1)} = \frac{1-v^2}{2}\,\varepsilon_\mathrm{max}^{(1)} = \frac{8}{\cosh^2 R}
\end{equation}
at $R\le\mbox{arsinh}\,1$, and
\begin{equation}
p_\mathrm{max}^{(1)} = (1-v^2)\,u_\mathrm{max}^{(1)} = \frac{1-v^2}{v^2}\,k_\mathrm{max}^{(1)} = \frac{1-v^2}{2}\,\varepsilon_\mathrm{max}^{(1)} = 2\coth^2 R
\end{equation}
at $R\ge\mbox{arsinh}\,1$.


\section{Multi-kink scattering}\label{sec:scattering}

We studied the collisions of two kinks (a kink and an antikink), three kinks (two kinks and an antikink), and four kinks (two kinks and two antikinks) at one point. Our main goal was to obtain $R$-dependences of the maximal values of the energy densities --- kinetic, gradient, potential and total. To do this, we numerically solved the discretized version of the equation of motion \eqref{eq:eqmo},
\begin{eqnarray}\label{discrete}
\frac{d^2\phi_n}{dt^2} &-& \frac{1}{h^2}(\phi_{n-1} -2\phi_{n}+\phi_{n+1}) 
+\displaystyle\frac{1}{12h^2}(\phi_{n-2}-4\phi_{n-1}+6\phi_{n}-4\phi_{n+1}+\phi_{n+2})\nonumber\\
&+&\tanh^2R \:\sin \phi_{n}-\frac{2}{\cosh^2R}\sin\frac{\phi_{n}}{2}  = 0,
\end{eqnarray}
using St\"ormer method of integration. Here $h=0.025$ is the space step, $n = 0, \pm 1, \pm 2,...$ and $\phi_{n}(t) = \phi(nh,t)$. The second derivative of the field $\phi$ with respect to $x$, $\phi_{xx}$, is discretized with the accuracy $O(h^4)$. In our numerical experiments we used the range $-375\le x\le 375$. Integration with respect to time was performed with the step $\tau=0.005$. We used $0\le t\le 500$ for searching maximal energy densities over $x$ and $t$.

The initial conditions for each case (collisions of two, three and four kinks) will be written below. In the case of collisions of three and four kinks, the initial conditions were fit in such a way that the kinks collided in one point.

From the point of view of physical applications, the energy distribution in collision of two or several kinks is of great importance. The point is that a kink is a topological defect, which describes, for example, a planar domain wall or another formation (``defect'') in a real physical system. The kinks collision thus represents mathematical model of collision of such defects, and the energy densities show space-time distribution of energy.


\subsection{Kink-antikink scattering}\label{sec:2kinks}

First of all, note that collisions of the DSG kink and antikink have been studied in detail in recent paper \cite{Gani.EPJC.2018.dsg}. An interesting phenomenon was found --- oscillons in the final state. Besides, the dependence of the critical velocity $v_\mathrm{cr}^{}$ on $R$ obtained in \cite{Gani.EPJC.2018.dsg} was somewhat different from that presented in classical paper \cite{Campbell.PhysD.1986.dsg}. We think that a series of local maxima of the dependence $v_\mathrm{cr}^{}(R)$ observed in \cite{Gani.EPJC.2018.dsg} is a consequence of more detailed investigation of appropriate range of the initial velocities of the colliding kinks.

\begin{figure*}[t!]
\begin{center}
  \centering
  \subfigure[\:kink+antikink]{\includegraphics[width=0.325
 \textwidth]{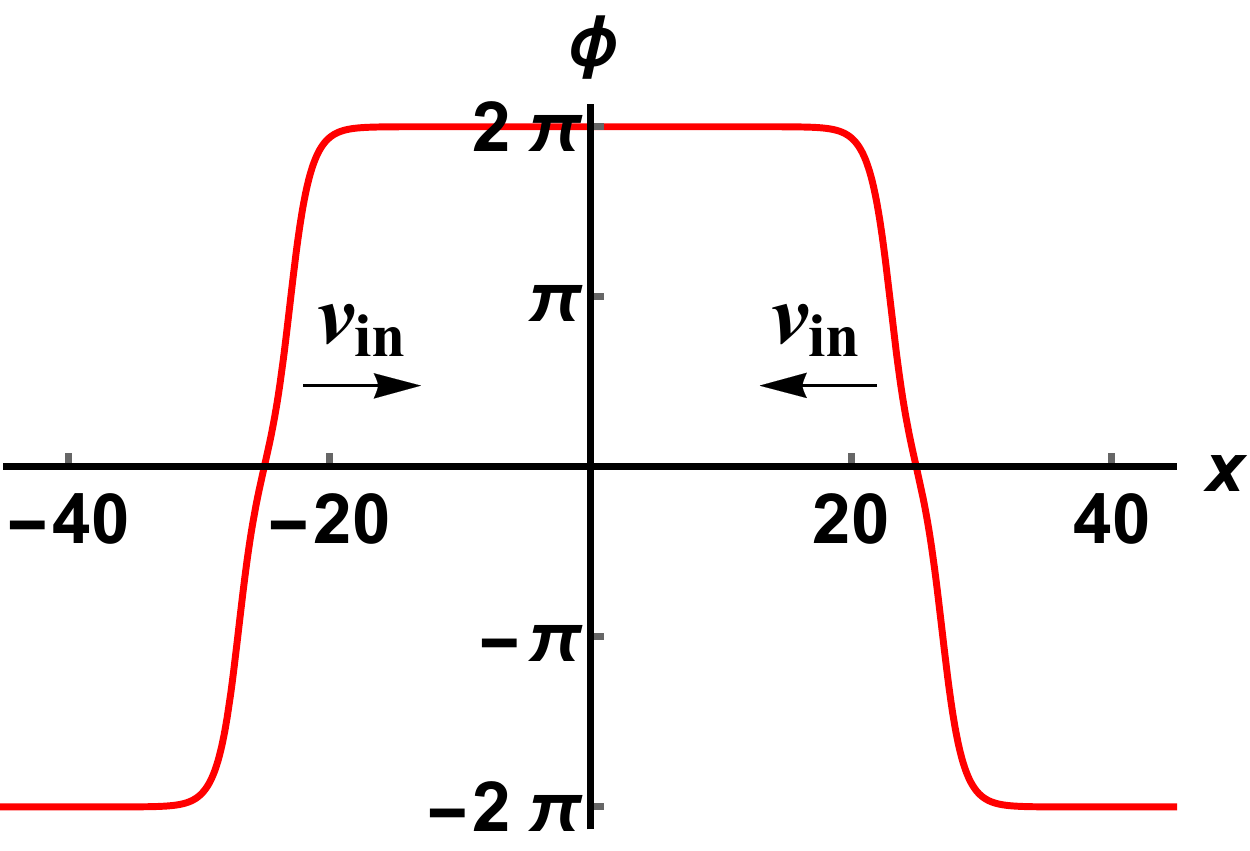}\label{fig:kak}}
  \subfigure[\:kink+antikink+kink]{\includegraphics[width=0.325
 \textwidth]{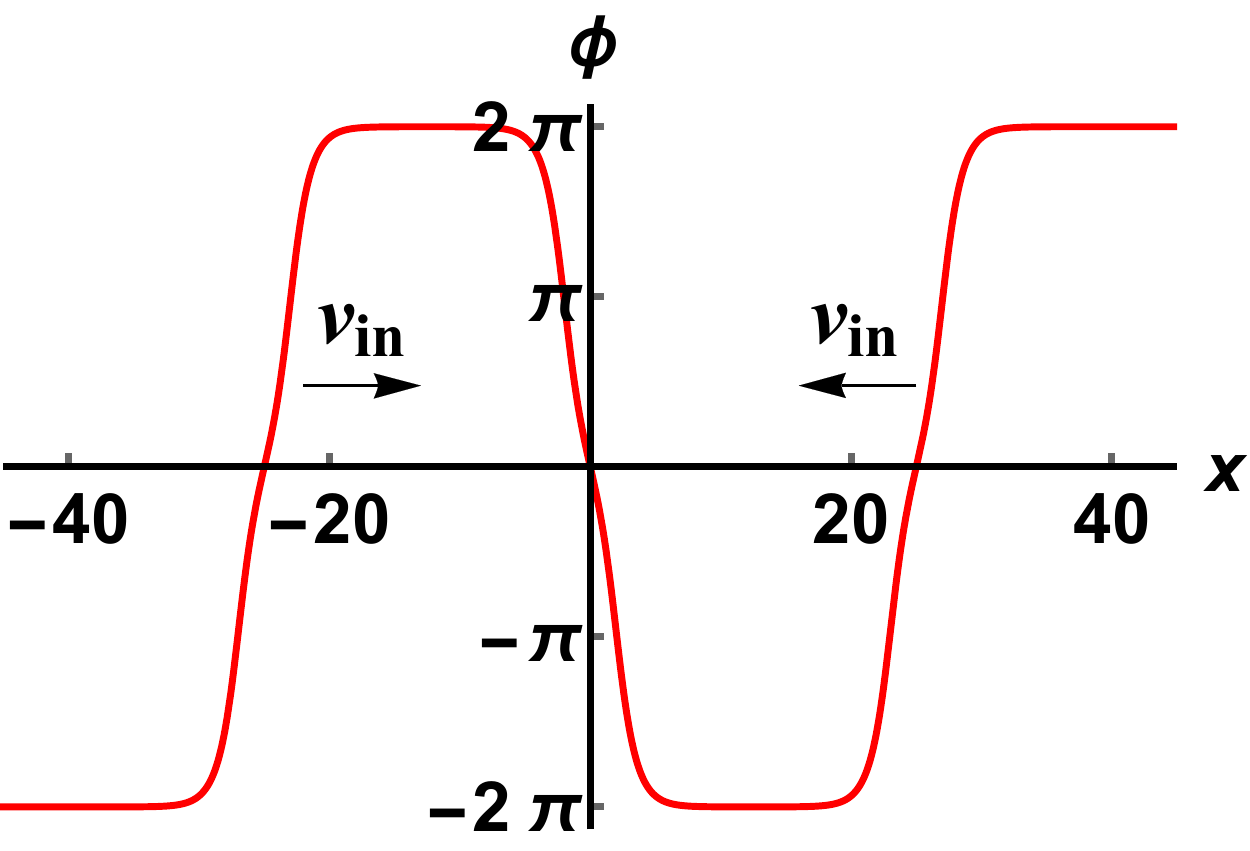}\label{fig:kakk}}
  \subfigure[\:kink+antikink+kink+antikink]{\includegraphics[width=0.325
 \textwidth]{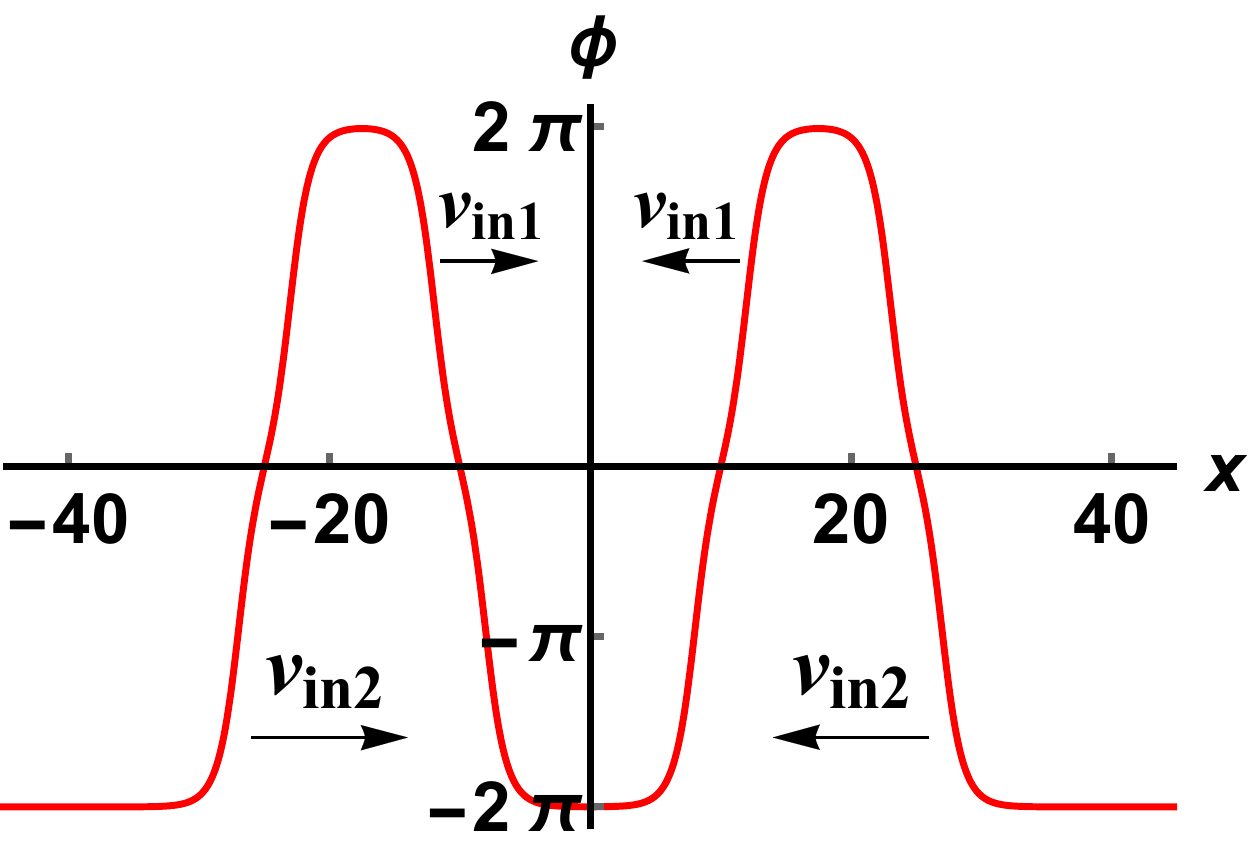}\label{fig:kakkak}}
  \\
  \caption{The initial configurations for $R=2$: (a) Eq.~\eqref{eq:DSG_2kinks}, (b) Eq.~\eqref{eq:DSG_3kinks}, (c) Eq.~\eqref{eq:DSG_4kinks}.}
  \label{fig:Potential}
\end{center}
\end{figure*}

\begin{figure*}[t!]
\begin{center}
  \centering
  \subfigure[\:$R=0.5$]{\includegraphics[width=0.3
 \textwidth]{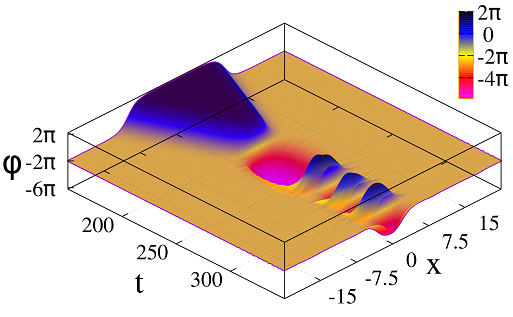}\label{fig:F2KR050V01X25}}
  \subfigure[\:$R=1.0$]{\includegraphics[width=0.3
 \textwidth]{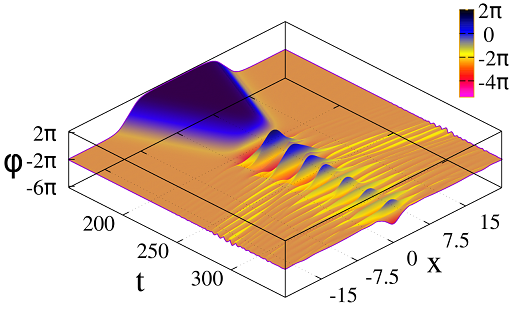}\label{fig:F2KR100V01X25}}
  \subfigure[\:$R=1.5$]{\includegraphics[width=0.3
 \textwidth]{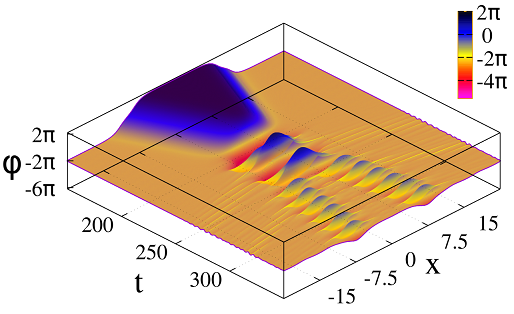}\label{fig:F2KR150V01X25}}
 \\
  \subfigure[\:$R=2.0$]{\includegraphics[width=0.3
 \textwidth]{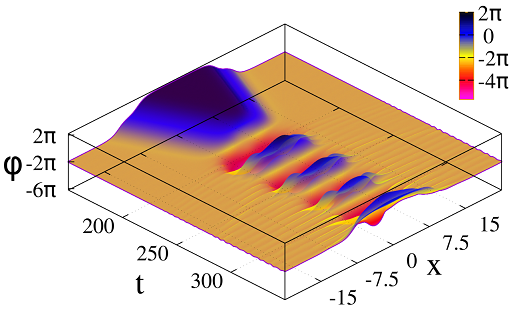}\label{fig:F2KR200V01X25}}
  \subfigure[\:$R=2.5$]{\includegraphics[width=0.3
 \textwidth]{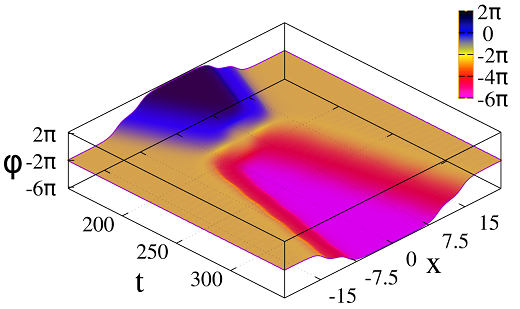}\label{fig:F2KR250V01X25}}
  \subfigure[\:$R=3.0$  ]
{\includegraphics[width=0.3\textwidth]{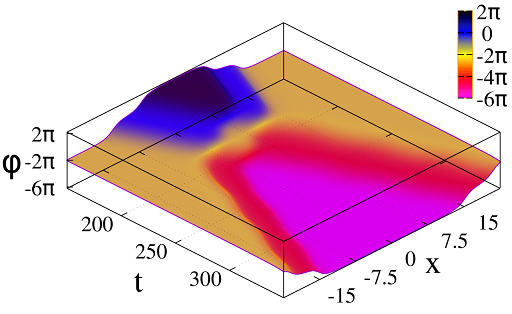}\label{fig:F2KR300V01X25}
}
 \\
 \caption{Space-time picture of the kink-antikink collisions for different $R$'s.}
  \label{fig:Fields1}
\end{center}
\end{figure*}

Unlike the previous works \cite{Gani.EPJC.2018.dsg,Gani.PRE.1999,Campbell.PhysD.1986.dsg}, we focus on study of maximal energy densities in collisions of the kink and antikink at the initial velocity $v_\mathrm{in}^{}=0.1$ and various $R$. To do this we used the initial condition in the form of the kink and antikink, which are initially placed at $x=-x_0$ and $x=+x_0$ and moving towards each other with the velocities $v=+v_\mathrm{in}^{}$ and $v=-v_\mathrm{in}^{}$, respectively,
\begin{equation}\label{eq:DSG_2kinks}
\phi_{\mathrm{k}\mathrm{\bar k}}(x) = \phi_{\mathrm{k}}\left(\frac{x+x_0^{}-v_\mathrm{in}^{} t}{\sqrt{1-v_\mathrm{in}^{2}}}\right) + \phi_{\mathrm{\bar k}}\left( \frac{x-x_0^{}+v_\mathrm{in}^{} t}{\sqrt{1-v_\mathrm{in}^{2}}}\right) - 2\pi,
\end{equation}
where $\phi_{\mathrm{k}(\mathrm{\bar k})}(x)$ is defined by Eq.~\eqref{eq:DSG_kinks_1} with $n=0$. This initial configuration is shown in Fig.~\ref{fig:kak}.
In our simulations we used $x_0=25$ and $v_\mathrm{in}^{}=0.1$. Space-time picture of the collision for some selected values of $R$ is shown in Fig.~\ref{fig:Fields1}.

\begin{figure}[t!]
\centering
 \subfigure[\:total]{\includegraphics[width=0.24
 \textwidth]{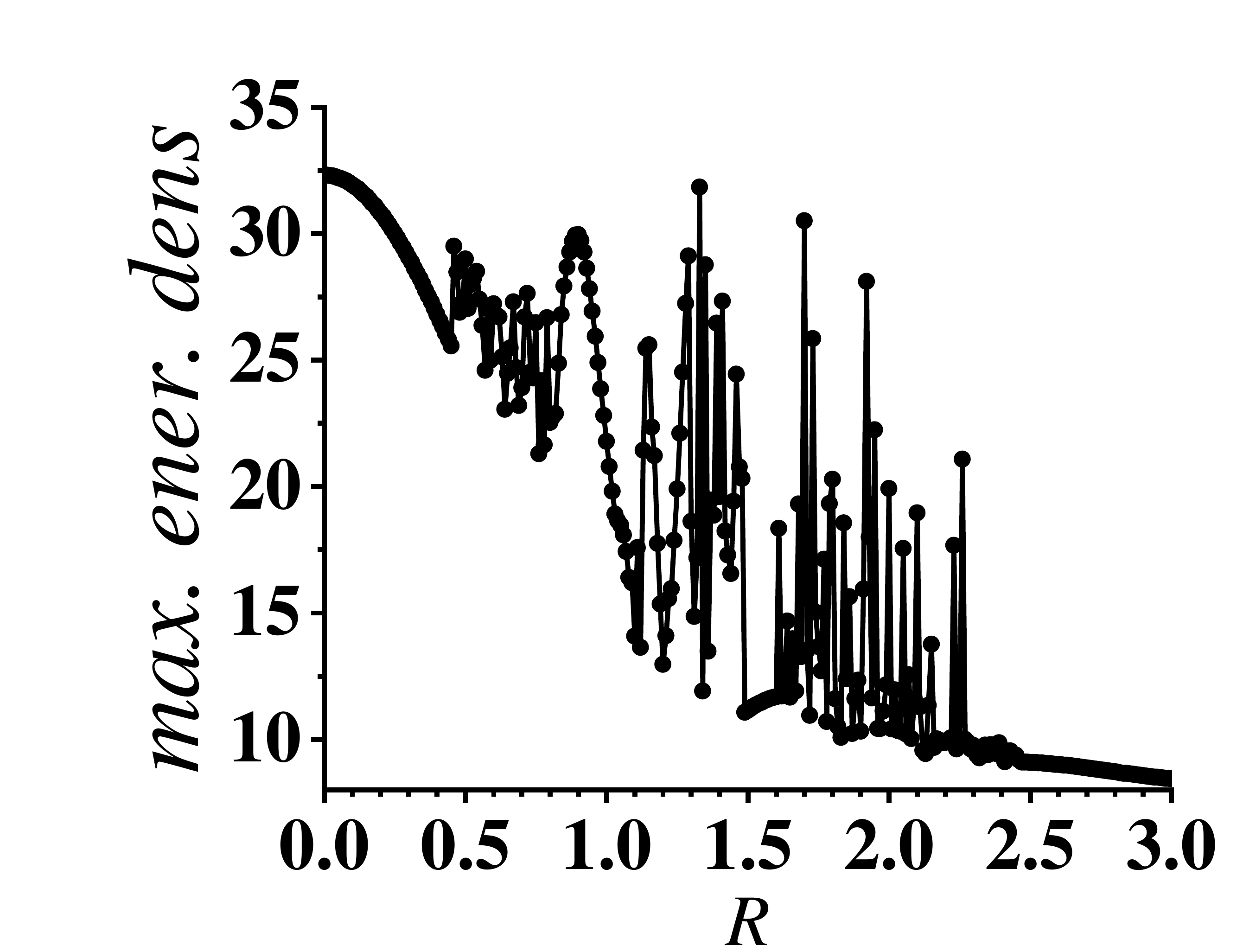}\label{fig:2kinksMaximumEnergyDensityV010}}
  \subfigure[\:kinetic]{\includegraphics[width=0.24
 \textwidth]{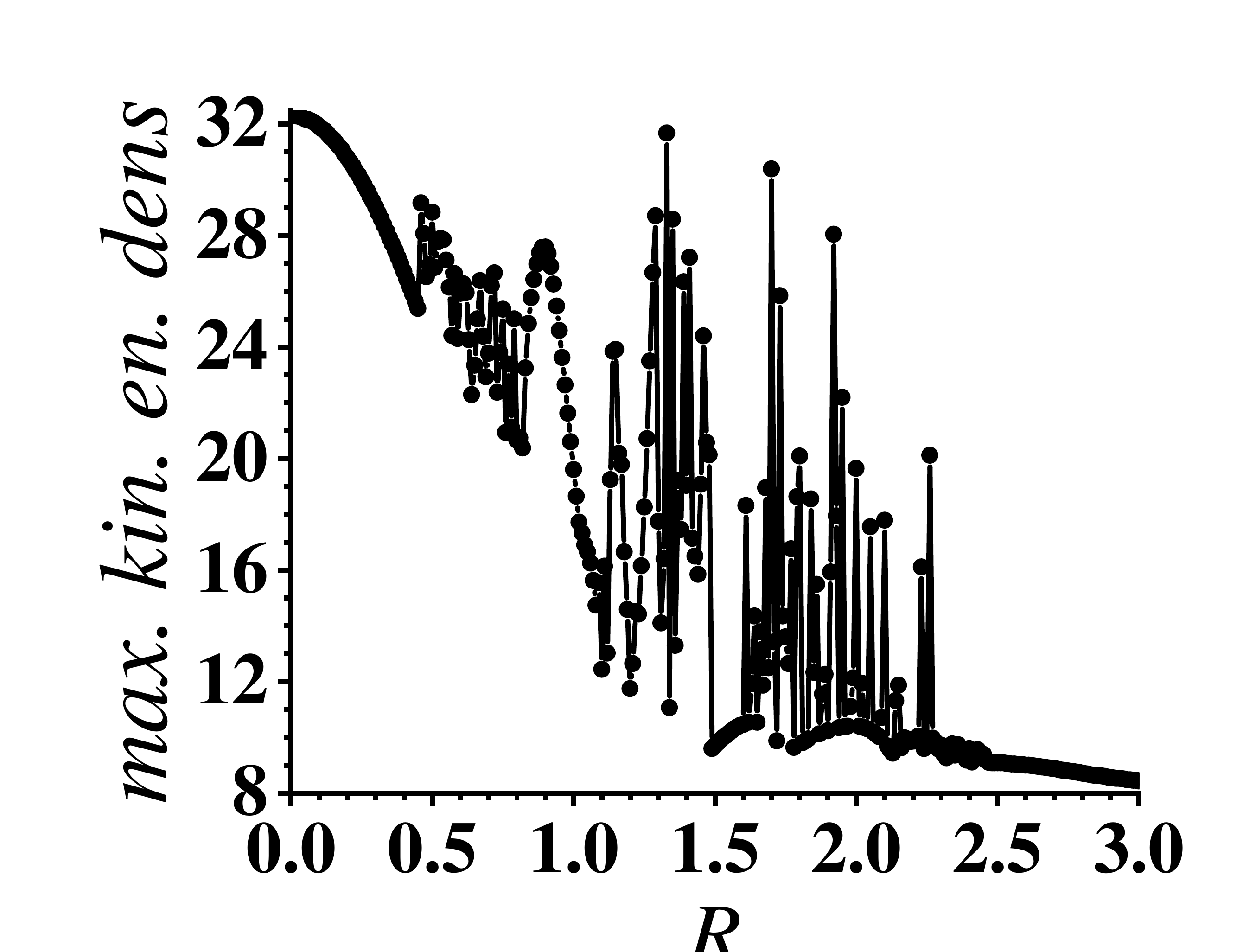}\label{fig:2kinksMaximumKineticEnergyDensityV010}}
      \subfigure[\:potential]{\includegraphics[width=0.24
 \textwidth]{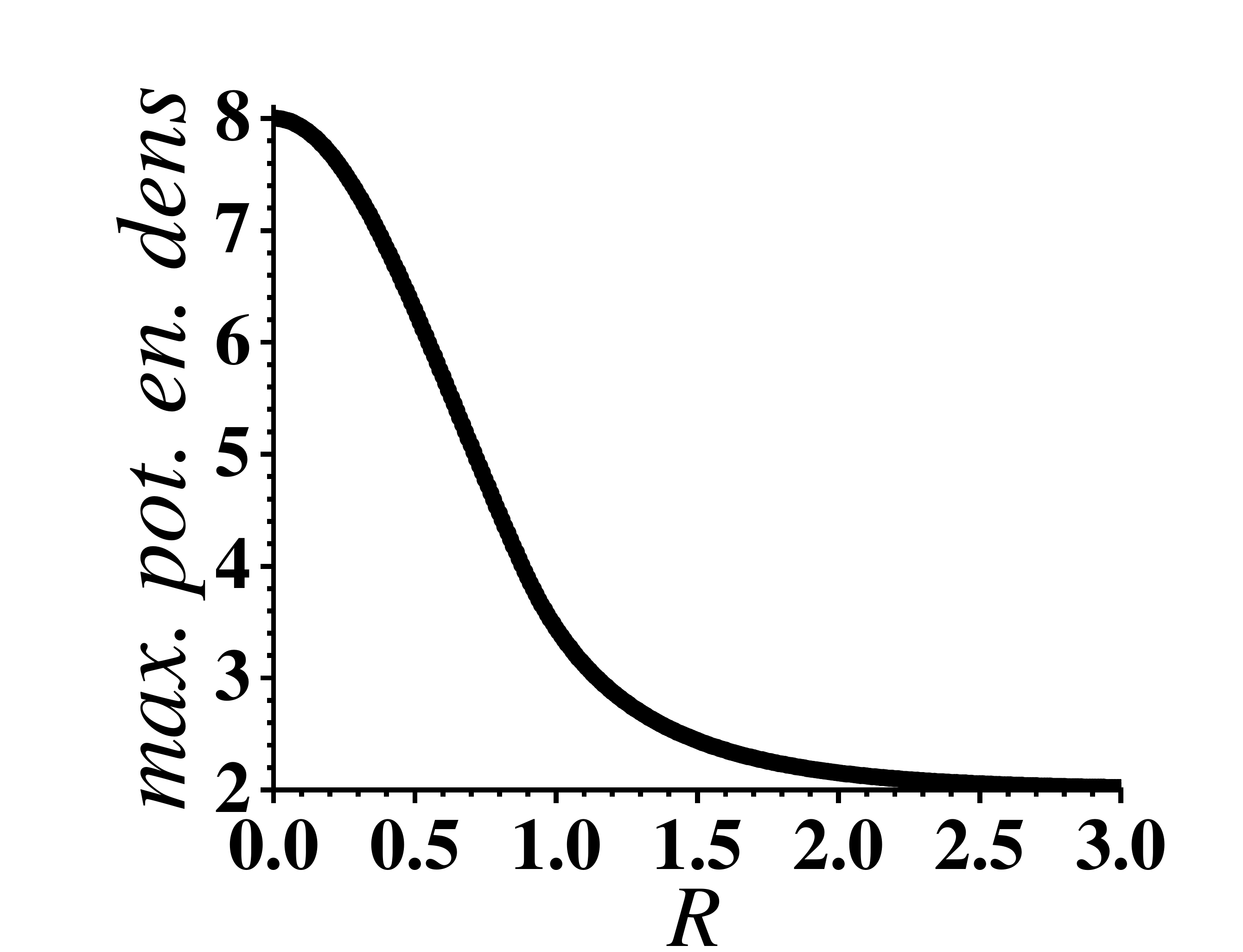}\label{fig:2kinksMaximumPotentialEnergyDensityV010}}
  \subfigure[\:gradient]{\includegraphics[width=0.24
 \textwidth]{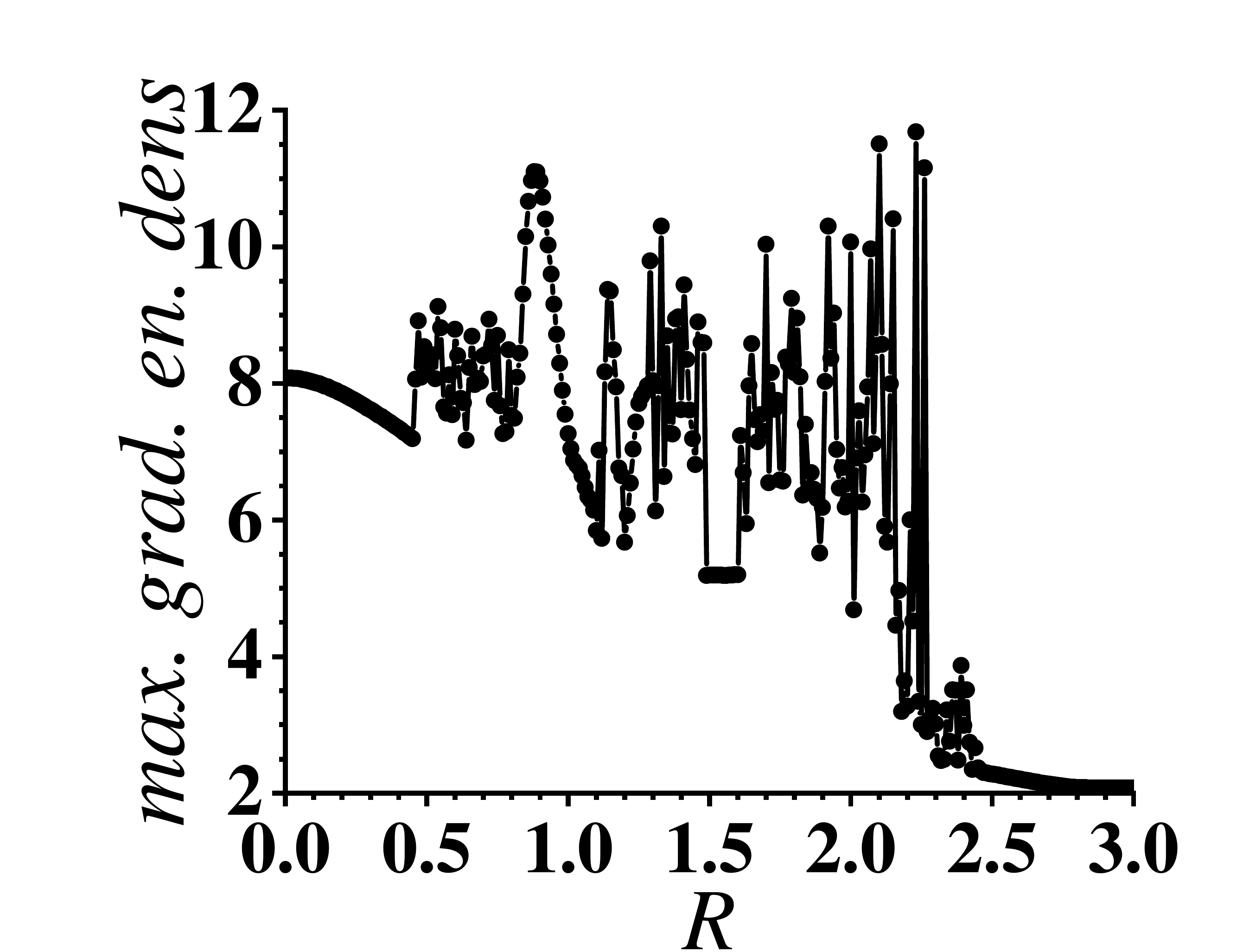}\label{fig:2kinksMaximumGradientEnergyDensityV010}}
\caption{Maximal energy densities as functions of $R$ for kink-antikink collisions at $v_\mathrm{in}^{}=0.1$}. 
\label{fig:2kinksMaxEnergyDensites}
\end{figure}

We performed numerical simulation of the kink-antikink collisions within a wide range of the parameter $R$. We have obtained the dependences of the maximal energy densities on the parameter $R$, see Fig.~\ref{fig:2kinksMaxEnergyDensites}.
These dependences are rather complicated. At $R\lsim 0.5$ and $R\gsim 2.5$ we see smooth and monotonously decreasing curves. At the same time, within the range $0.5\lsim R\lsim 2.5$ the dependences look quite stochastic. The observed difference in behavior of the maximal energy densities at various $R$ could be a consequence of different character of kink-antikink interaction at given initial conditions. In particular, depending on the parameter $R$, the initial velocity $v_\mathrm{in}^{}=0.1$ can be either less or more than the critical value $v_\mathrm{cr}^{}$. The point is that the critical velocity depends on $R$ in a complicated way \cite{Gani.EPJC.2018.dsg}.

Analyzing the dependence $v_\mathrm{cr}^{}(R)$ which is presented in Fig.~6 of Ref.~\cite{Gani.EPJC.2018.dsg}, we can see that the initial velocity $v_\mathrm{in}^{}=0.1$ is less than the critical in the range $0.5\lsim R\lsim 2.5$. Thus complicated and almost stochastic behavior of the maximal energy densities is observed for those $R$'s at which the kinks form a bound state, {\it i.e.}, a bion. On the other hand, at those $R$'s at which the kinks escape from each other after a collision, the smooth and monotonous dependences are observed. We can assume that the complicated dependences of the maximal energy densities on $R$ can be a consequence of complex resonance processes in bion. As a result, energy is redistributed that leads to abrupt changes of the maximal energy densities.

\begin{figure}[t!]
\centering
 \subfigure[\:total]{\includegraphics[width=0.24
 \textwidth]{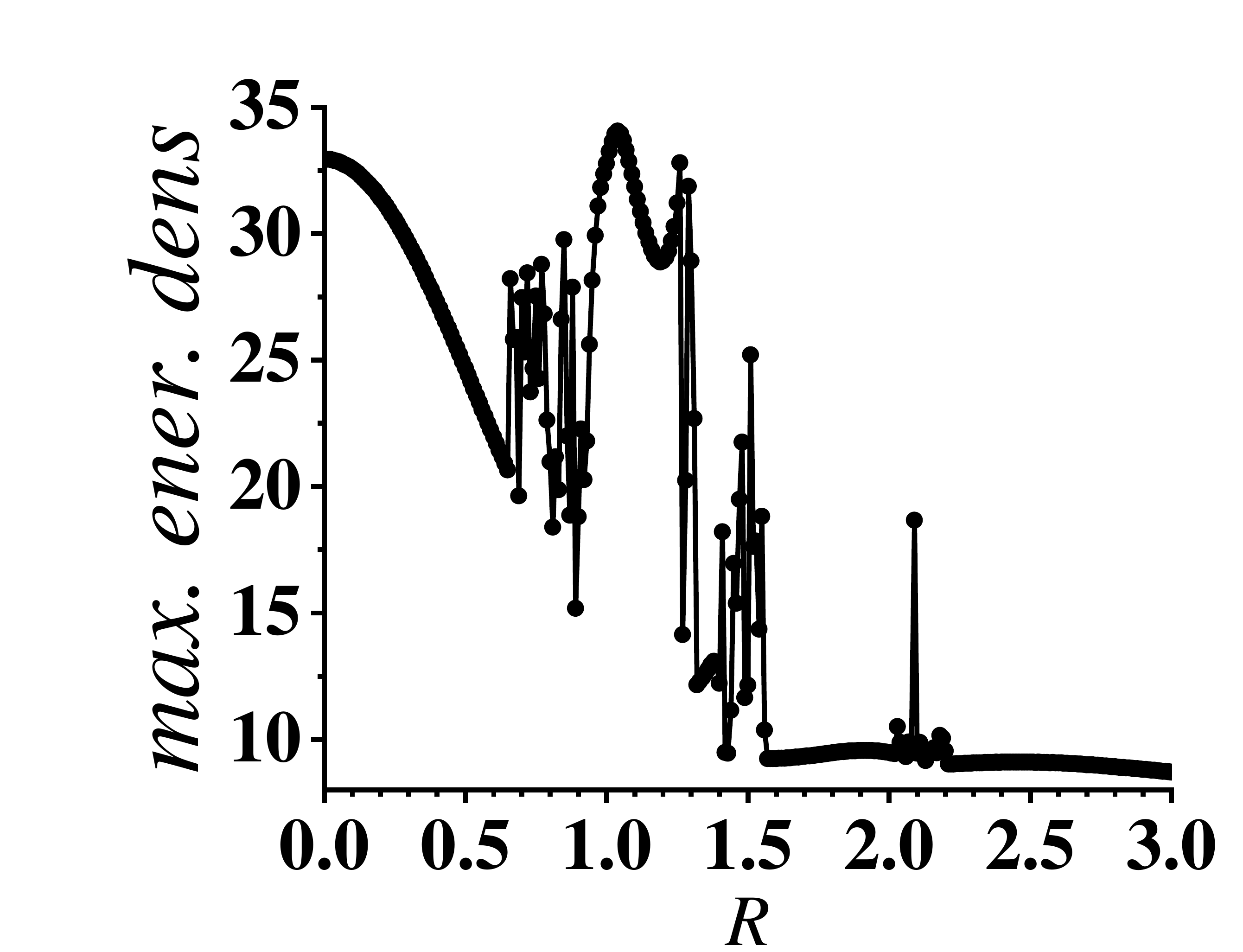}\label{fig:2kinksMaximumEnergyDensityV017}}
  \subfigure[\:kinetic]{\includegraphics[width=0.24
 \textwidth]{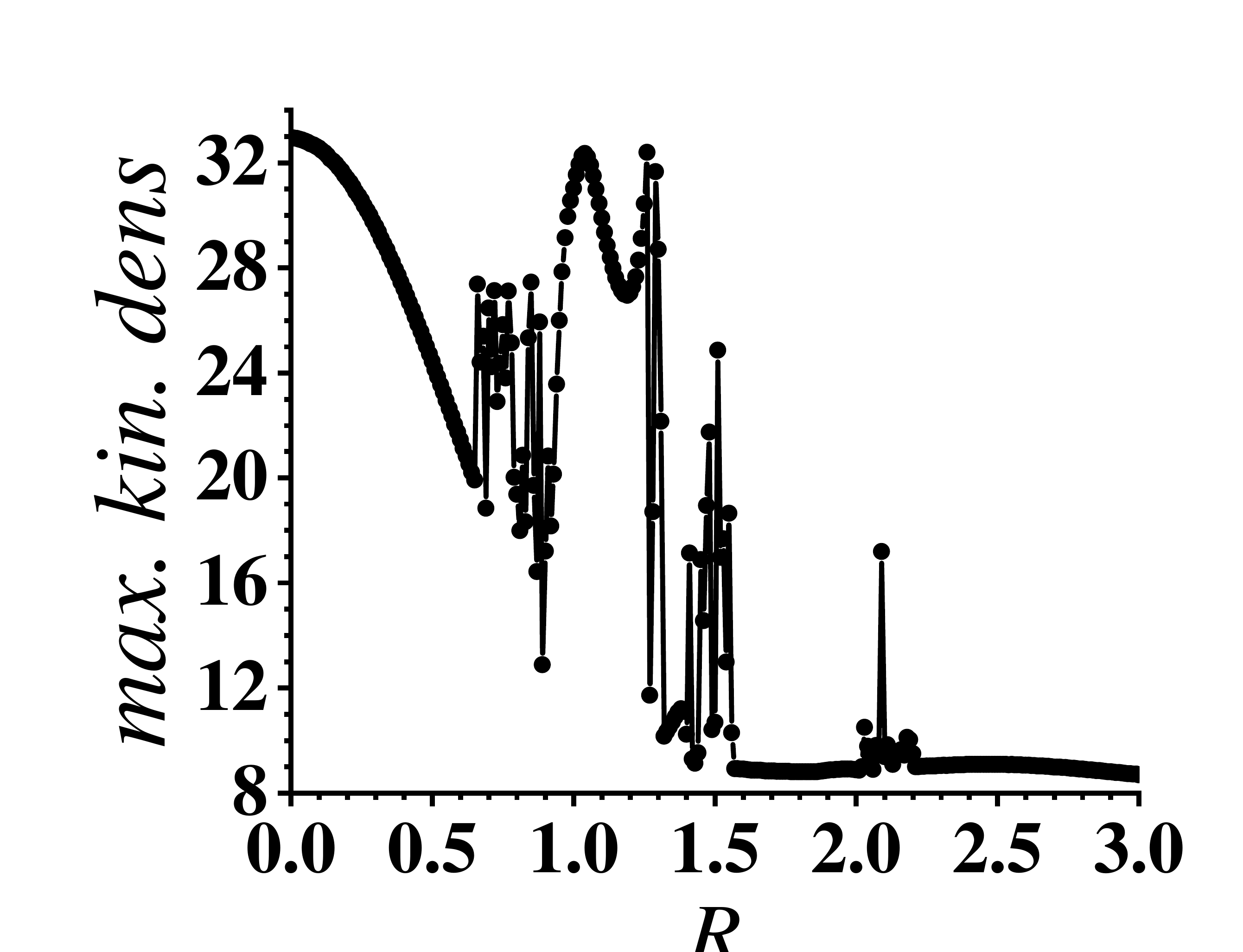}\label{fig:2kinksMaximumKineticEnergyDensityV017}}
      \subfigure[\:potential]{\includegraphics[width=0.24
 \textwidth]{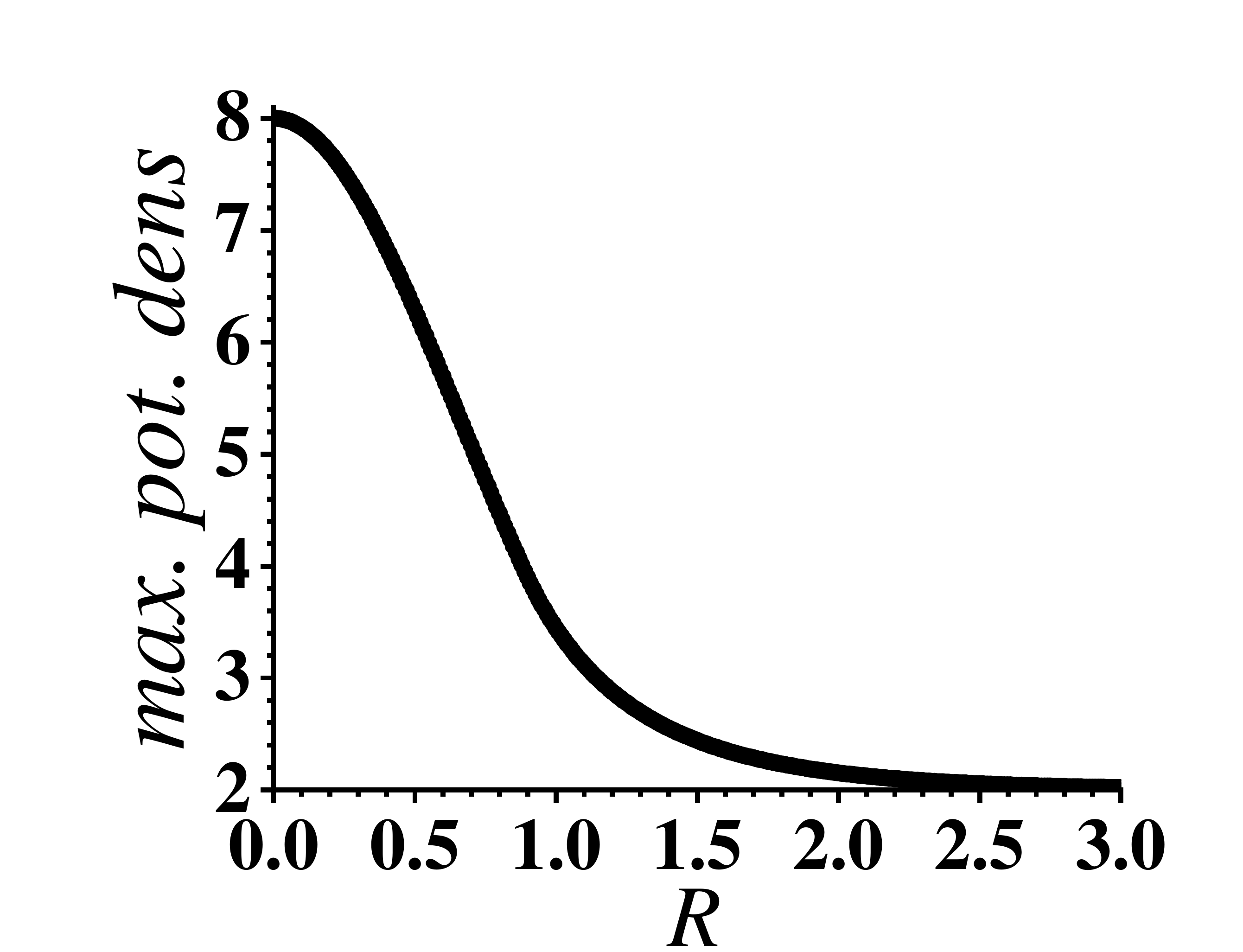}\label{fig:2kinksMaximumPotentialEnergyDensityV017}}
  \subfigure[\:gradient]{\includegraphics[width=0.24
 \textwidth]{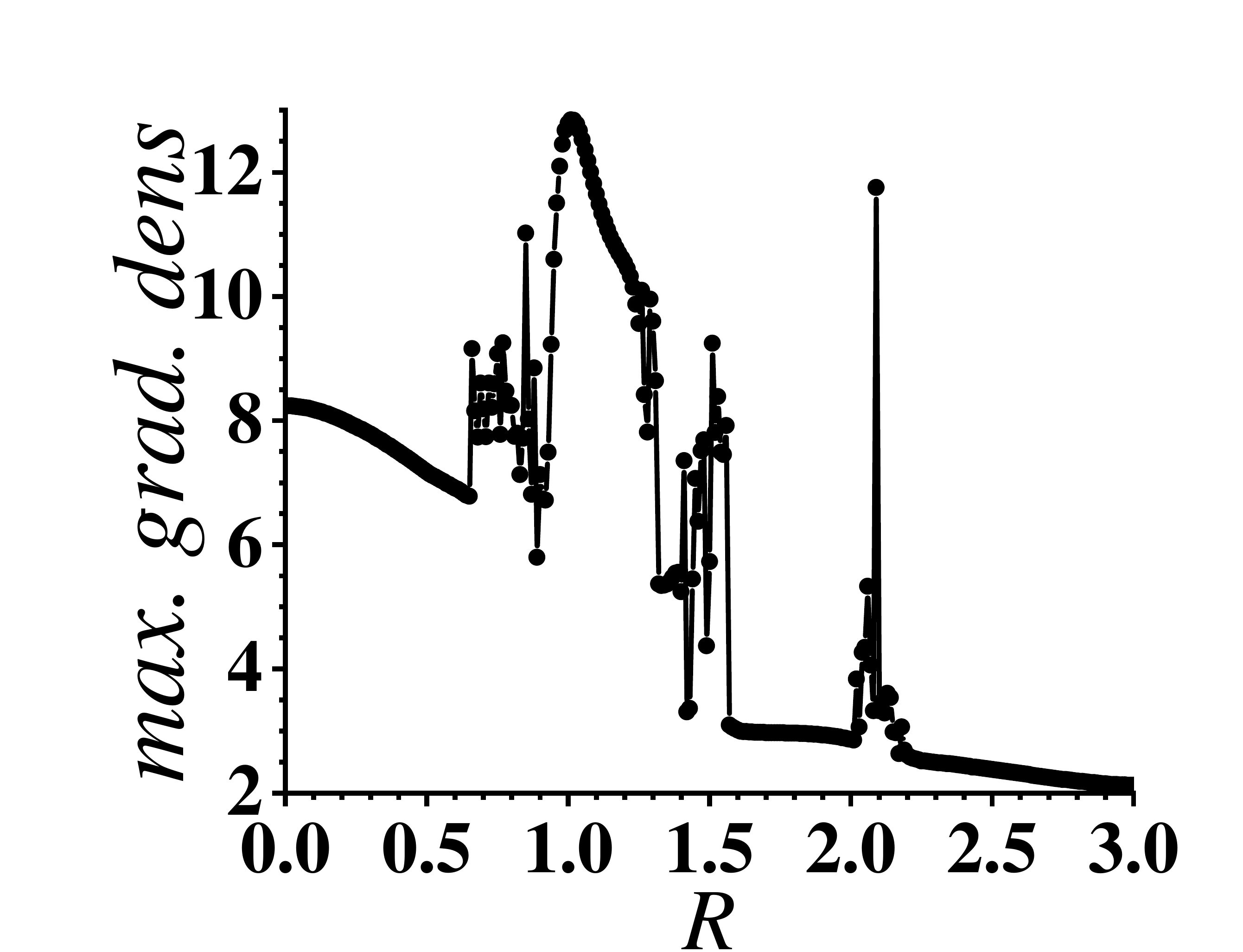}\label{fig:2kinksMaximumGradientEnergyDensityV017}}
\caption{Maximal energy densities as functions of $R$ for kink-antikink collisions at $v_\mathrm{in}^{}=0.17$.} 
\label{fig:2kinksMaxEnergyDensitesV017}
\end{figure}

\begin{figure}[t!]
\centering
 \subfigure[\:total]{\includegraphics[width=0.24
 \textwidth]{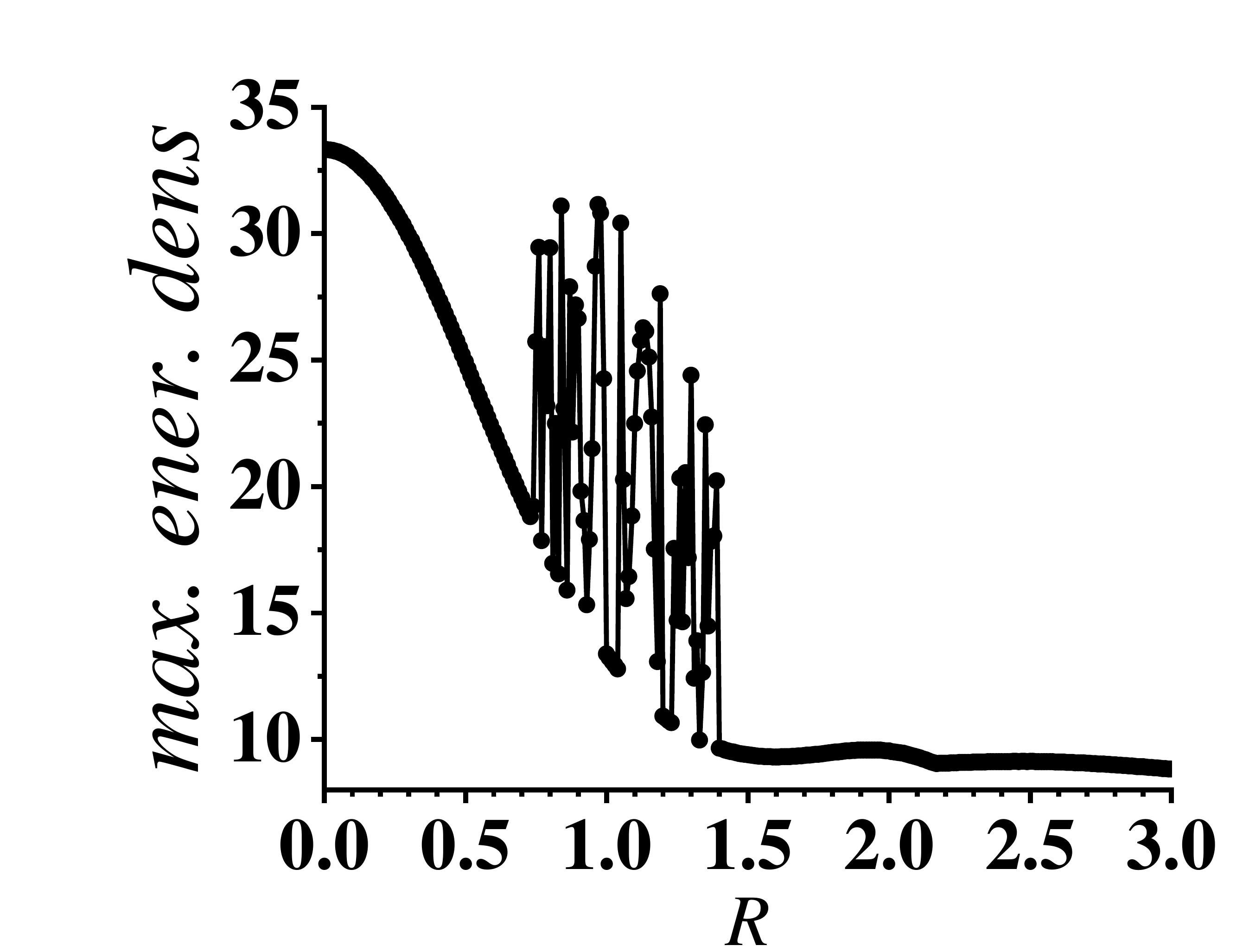}\label{fig:2kinksMaximumEnergyDensityV020}}
  \subfigure[\:kinetic]{\includegraphics[width=0.24
 \textwidth]{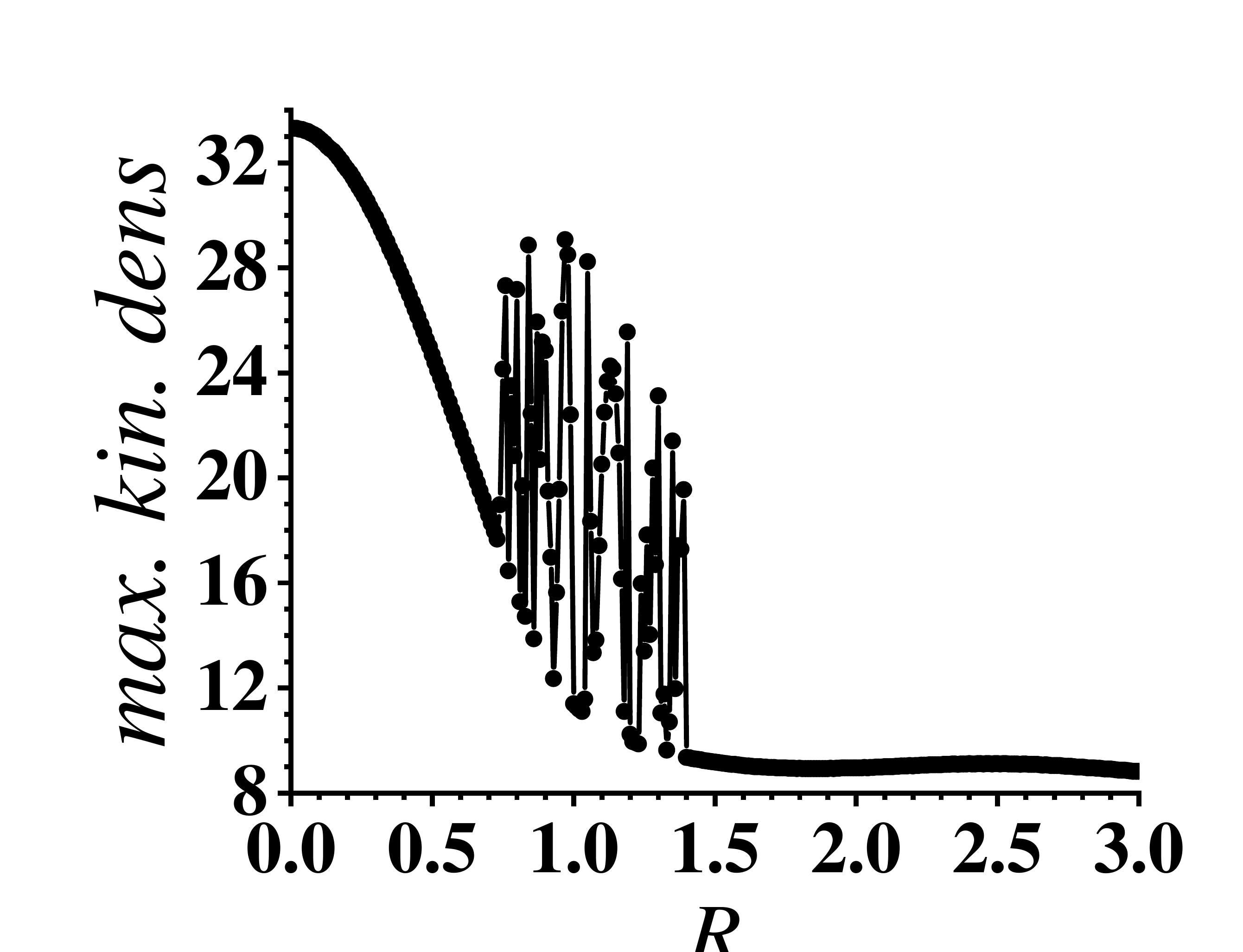}\label{fig:2kinksMaximumKineticEnergyDensityV020}}
      \subfigure[\:potential]{\includegraphics[width=0.24
 \textwidth]{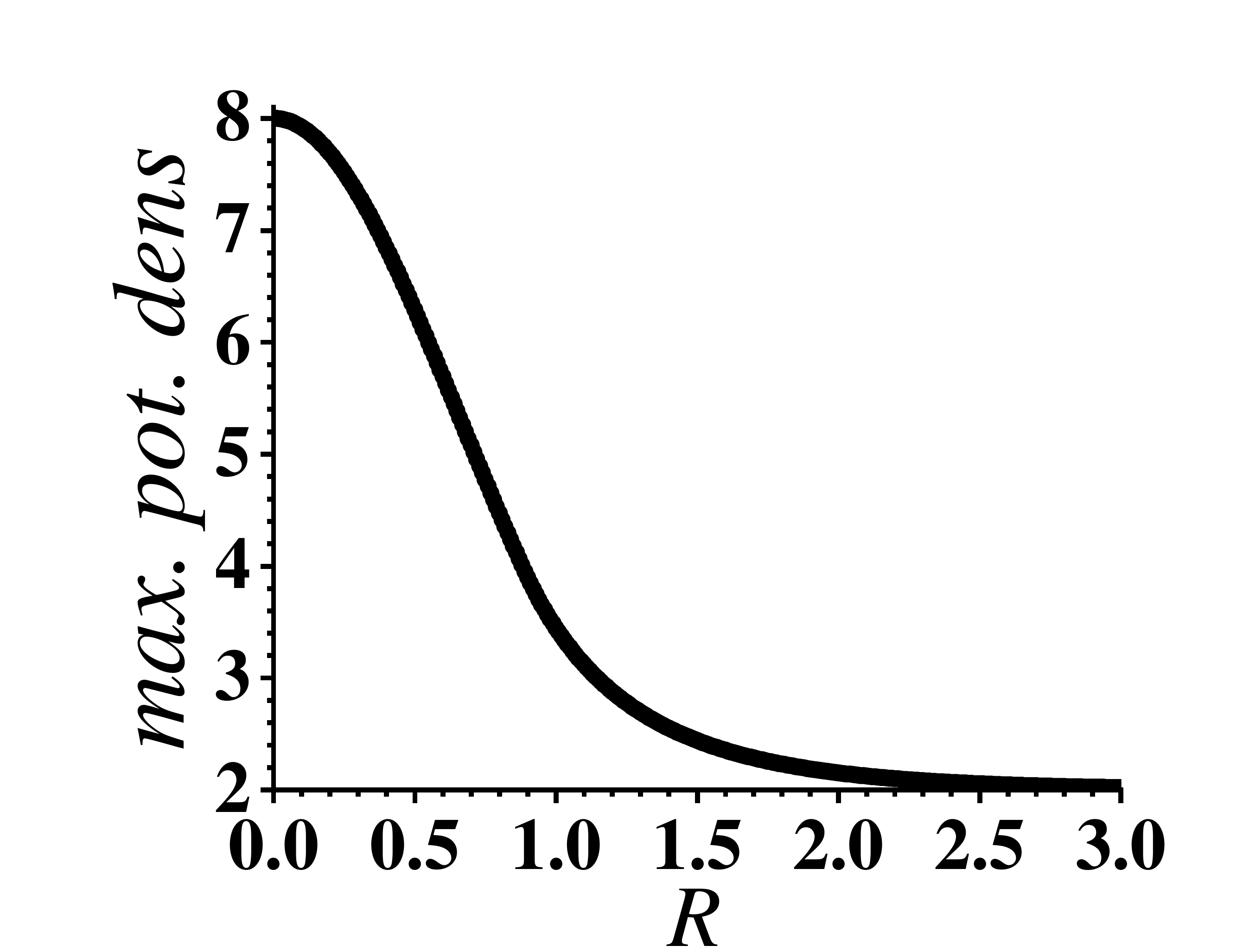}\label{fig:2kinksMaximumPotentialEnergyDensityV020}}
  \subfigure[\:gradient]{\includegraphics[width=0.24
 \textwidth]{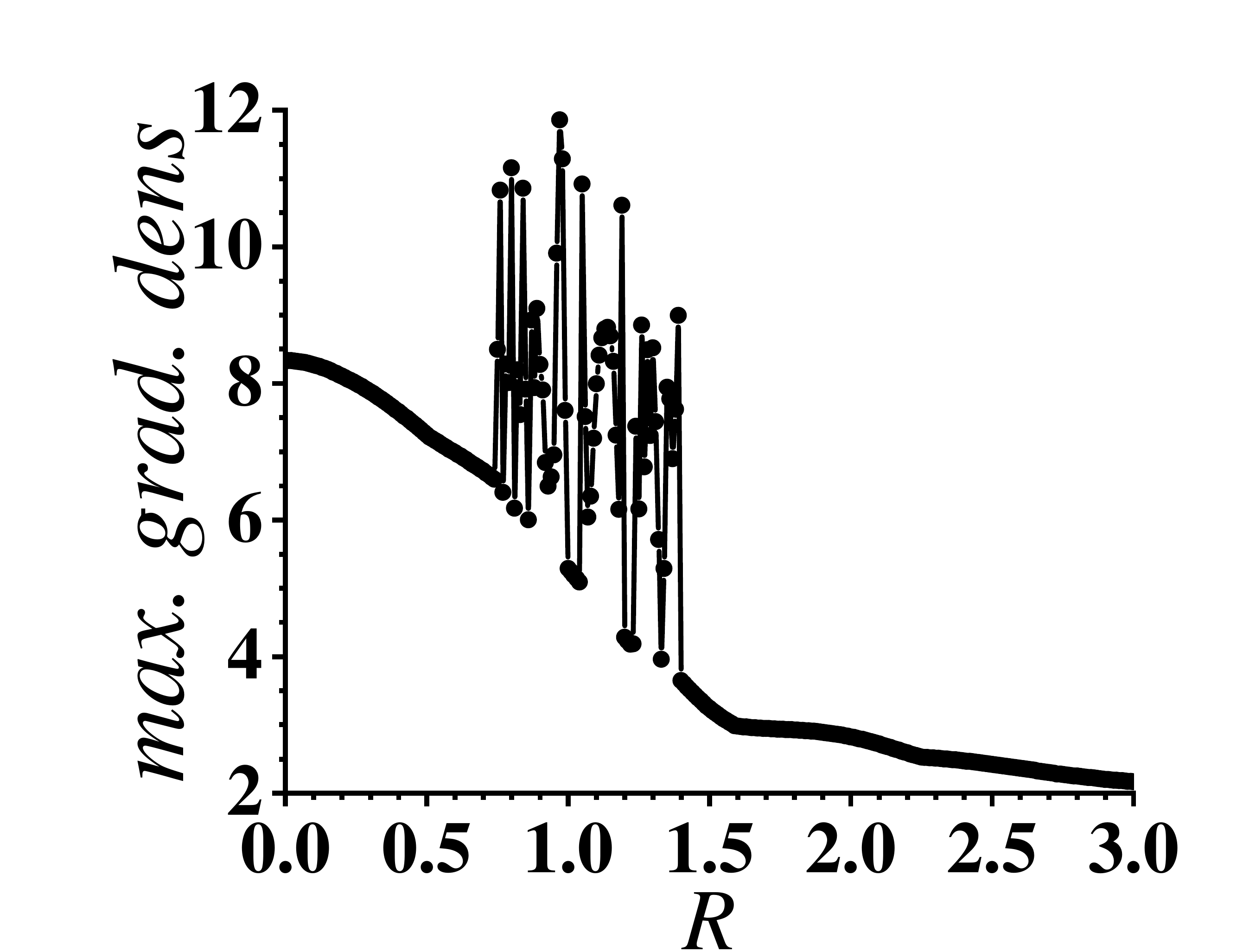}\label{fig:2kinksMaximumGradientEnergyDensityV020}}
\caption{Maximal energy densities as functions of $R$ for kink-antikink collisions at $v_\mathrm{in}^{}=0.2$.} 
\label{fig:2kinksMaxEnergyDensitesV020}
\end{figure}

\begin{figure}[t!]
\begin{center}
  \centering
   \subfigure[\:$R=1.49$]{\includegraphics[width=0.3
 \textwidth]{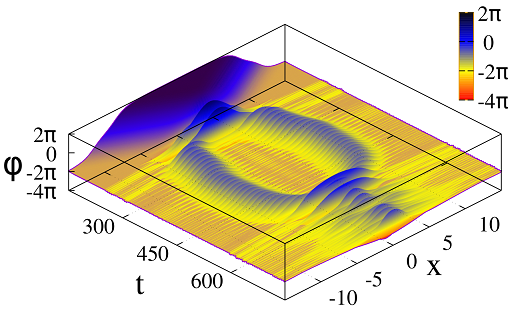}\label{fig:F2KR149V01X25}}
    \subfigure[\:$R=1.50$]{\includegraphics[width=0.3
 \textwidth]{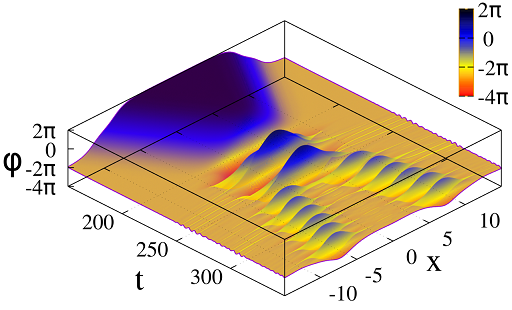}\label{fig:F2KR150V01X25N}}
  \subfigure[\:$R=1.52$]{\includegraphics[width=0.3
 \textwidth]{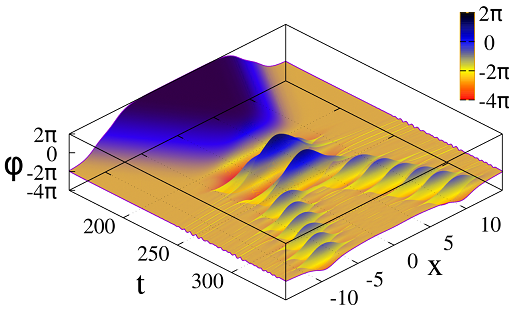}\label{fig:F2KR152V01X25}}
 \\
   \subfigure[\:$R=1.60$]{\includegraphics[width=0.3
 \textwidth]{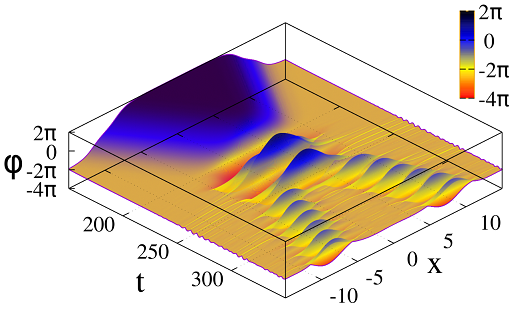}\label{fig:F2KR160V01X25}}
   \subfigure[\:$R=1.61$]{\includegraphics[width=0.3
 \textwidth]{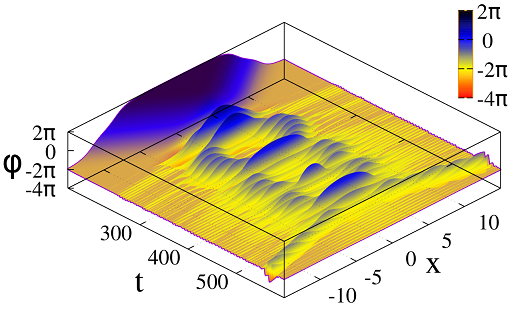}\label{fig:F2KR161V01X25}}
   \subfigure[\:$R=1.63$]{\includegraphics[width=0.3
 \textwidth]{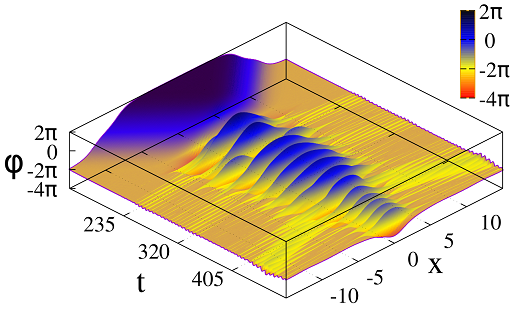}\label{fig:F2KR163V01X25}}
 \\
 \caption{Space-time picture of the kink-antikink collisions for some $R$'s from the range $1.49\le R\le 1.63$.}
  \label{fig:Fields2}
\end{center}
\end{figure}

In order to confirm our hypothesis, we performed numerical simulations of the kink-antikink collisions at $v_\mathrm{in}^{}=0.17$ and $v_\mathrm{in}^{}=0.2$. The maximal energy densities are shown in Figs.~\ref{fig:2kinksMaxEnergyDensitesV017} and \ref{fig:2kinksMaxEnergyDensitesV020}. From these figures one can see that regions of the stochastic behavior of the maximal energy densities become narrower. As it was for $v_\mathrm{in}^{}=0.1$, at these new initial velocities the stochastic behavior of the energy densities appears at those $R$'s at which $v_\mathrm{in}^{}<v_\mathrm{cr}^{}(R)$ and the kink and antikink capture each other and form a bion in the final state.
Another interesting fact is the presence of smooth and monotonous segments of the maximal energy densities in the range $1.5\lsim R\lsim 1.6$, see Fig.~\ref{fig:2kinksMaxEnergyDensites}. We can assume that it is a consequence of the fact that at these $R$'s the bound state of kinks evolves differently. The space-time picture of the kink-antikink collision at some selected values of the parameter $R$ from the range $1.49\le R\le 1.63$ is shown in Fig.~\ref{fig:Fields2}. It it clearly seen that at $1.5\lsim R\lsim 1.6$ two escaping oscillons in the final state are formed after two oscillations of the kink-antikink bound state. At other values of $R$ in Fig.~\ref{fig:Fields2} the oscillons in the final state are either formed after more complicated evolution of the kink-antikink bound state or not formed at all.
%


\subsection{Collision of two kinks and an antikink}\label{sec:3kinks}

\begin{figure}[t!]
\begin{center}
  \centering
     \subfigure[\:$R=0.50$]{\includegraphics[width=0.3
 \textwidth]{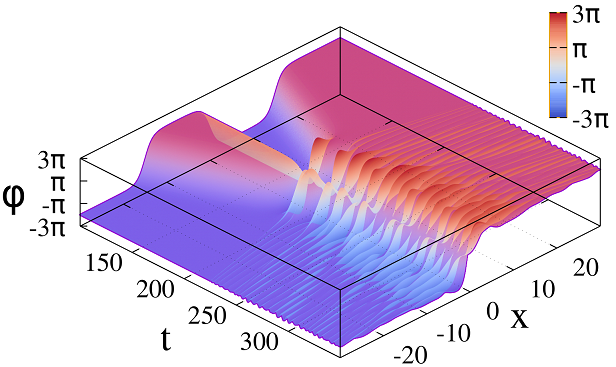}\label{fig:F3kR050V01X25}}
    \subfigure[\:$R=1.0$]{\includegraphics[width=0.3
 \textwidth]{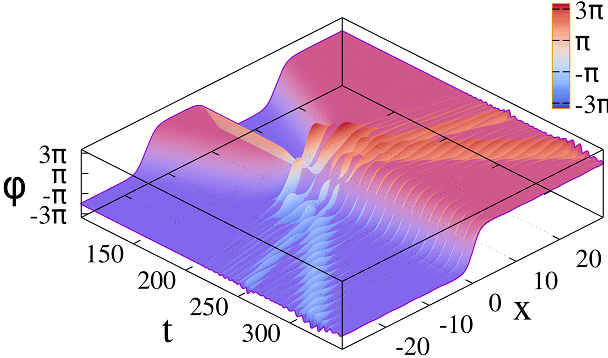}\label{fig:F3kR100V01X25}}
  \subfigure[\:$R=1.5$]{\includegraphics[width=0.3
 \textwidth]{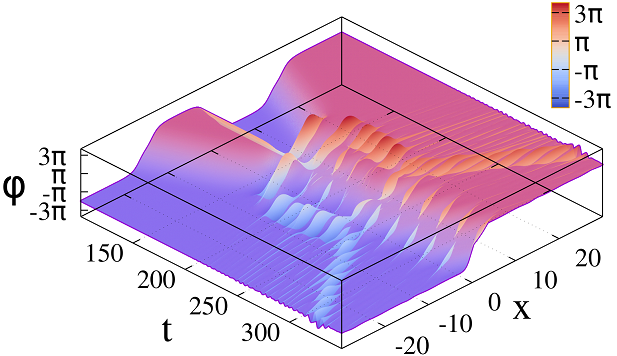}\label{fig:F3KR150V01X25}}
 \\
  \subfigure[\:$R=2.0$]
{\includegraphics[width=0.3\textwidth]{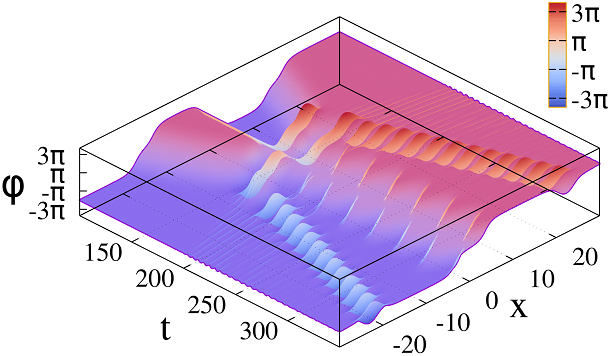}\label{fig:F3kR200V01X25}
}
   \subfigure[\:$R=2.5$]{\includegraphics[width=0.3
 \textwidth]{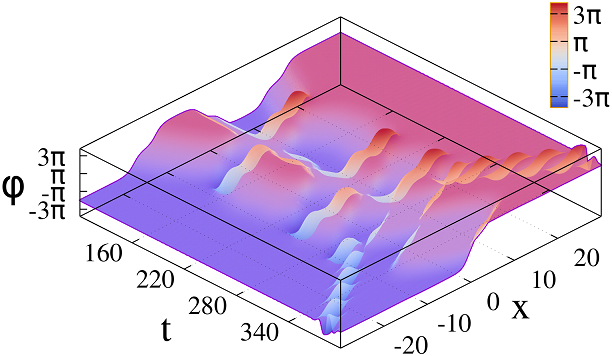}\label{fig:F3kR250V01X25}}
   \subfigure[\:$R=3.0$]{\includegraphics[width=0.3
 \textwidth]{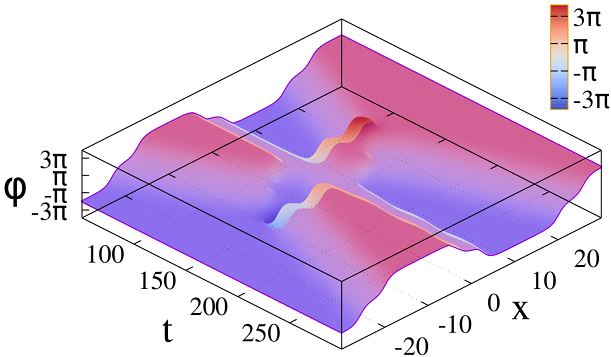}\label{fig:F3KR300V01X25}}
 \caption{Space-time picture of collisions of three kinks for different $R$'s.}
  \label{fig:3kFields1}
\end{center}
\end{figure}

We performed numerical simulation of the collision of three DSG kinks in one point. We used the following initial configuration:
\begin{equation}\label{eq:DSG_3kinks}
\phi_{\mathrm{k\bar k k}}(x) = \phi_{\mathrm{k}}\left(\frac{x+x_0^{}-v_\mathrm{in}^{} t}{\sqrt{1-v_\mathrm{in}^{2}}}\right) + \phi_{\mathrm{\bar k}}(x) + \phi_{\mathrm{k}}\left( \frac{x-x_0^{}+v_\mathrm{in}^{} t}{\sqrt{1-v_\mathrm{in}^{2}}}\right),
\end{equation}
where $\phi_{\mathrm{k}(\mathrm{\bar k})}(x)$ is defined by Eq.~\eqref{eq:DSG_kinks_1} with $n=0$. This initial condition corresponds to two kinks and an antikink. The antikink is at rest at the origin $x=0$, and two kinks are moving towards the antikink with the initial velocities $v=+v_\mathrm{in}^{}$ and $v=-v_\mathrm{in}^{}$, starting from $x=-x_0$ and $x=+x_0$, respectively. This initial configuration is shown in Fig.~\ref{fig:kakk}.
In our simulations we used $x_0=25$ and $v_\mathrm{in}^{}=0.1$. Space-time picture of the collision for some selected values of $R$ is shown in Fig.~\ref{fig:3kFields1}.
The interaction of three kinks looks rather complicated. In particular, in a pair ``kink+antikink''  resonance interaction can occur. This leads to appearance of various intermediate and final states, see Fig.~\ref{fig:3kFields1}. (Note that the term ``final state'' stands here for configuration observed after a long time after the collision. ``Long time'', in turn, means a period of time which is much longer than life-times of various bound states (of oscillons, of kink and oscillon, and so on), if such states do not have infinite life-times.)

\begin{figure}[t!]
\centering
 \subfigure[\:total]{\includegraphics[width=0.24
 \textwidth]{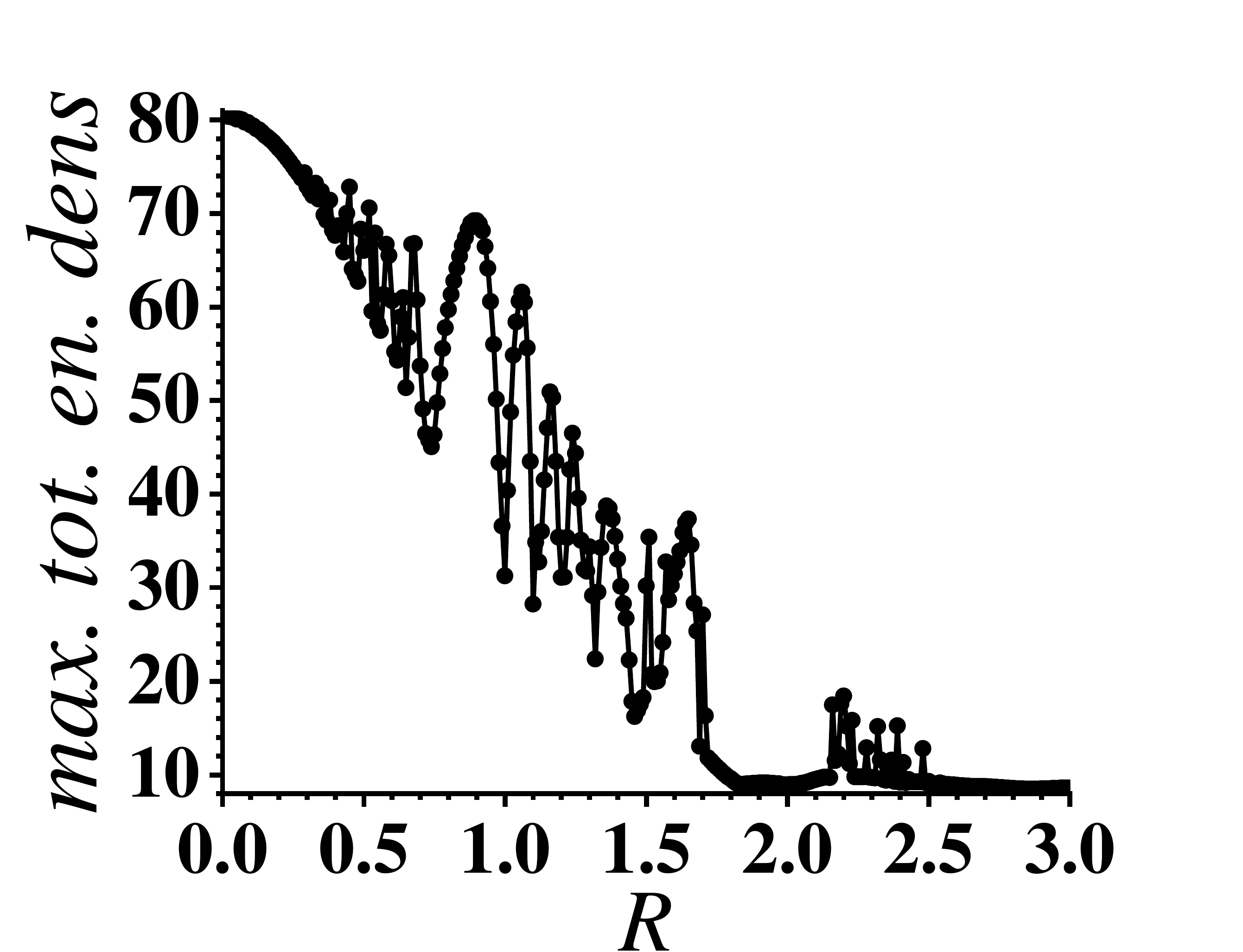}\label{fig:3kinksMaximumEnergyDensityV010}}
  \subfigure[\:kinetic]{\includegraphics[width=0.24
 \textwidth]{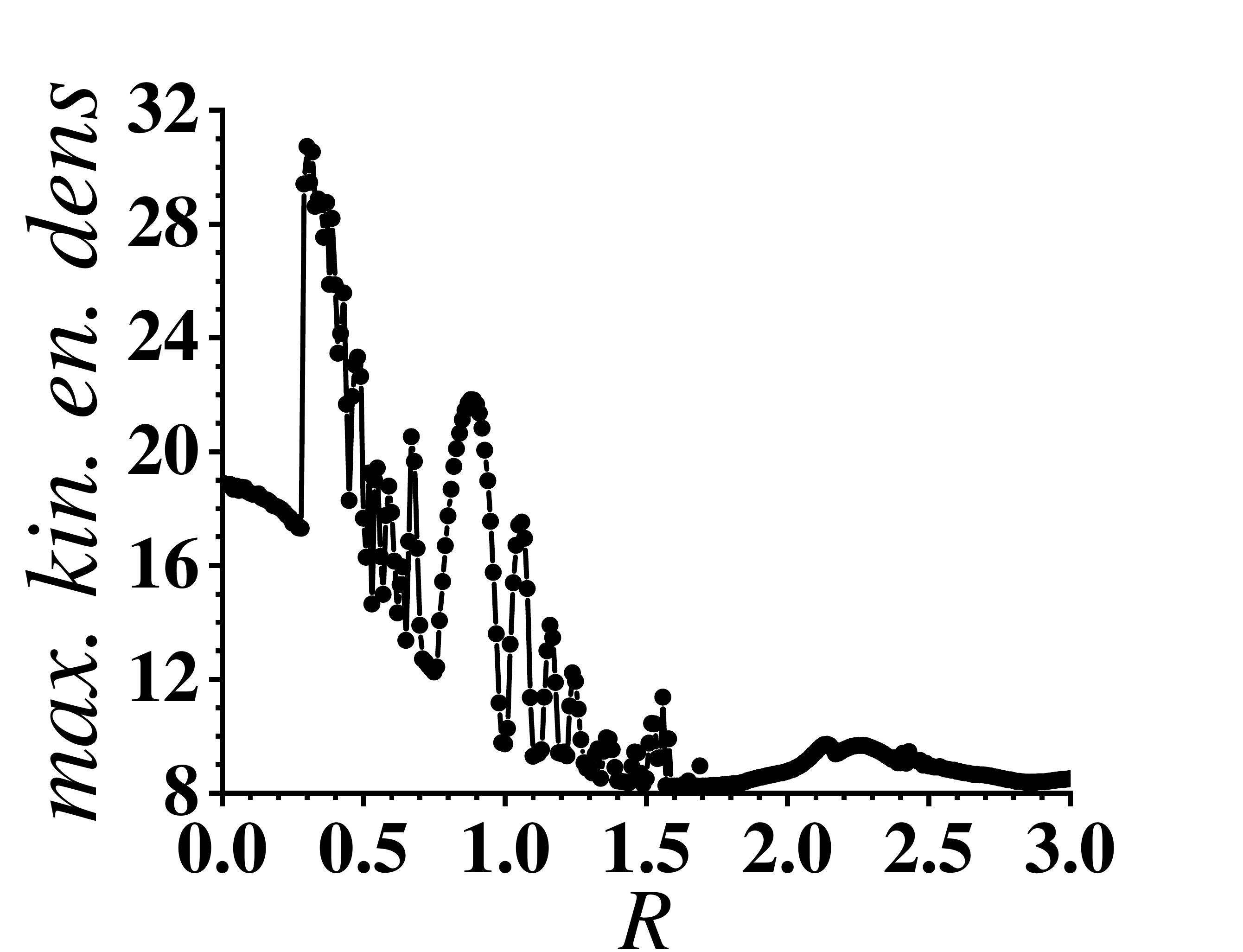}\label{fig:3kinksMaximumKineticEnergyDensityV010}}
      \subfigure[\:potential]{\includegraphics[width=0.24
 \textwidth]{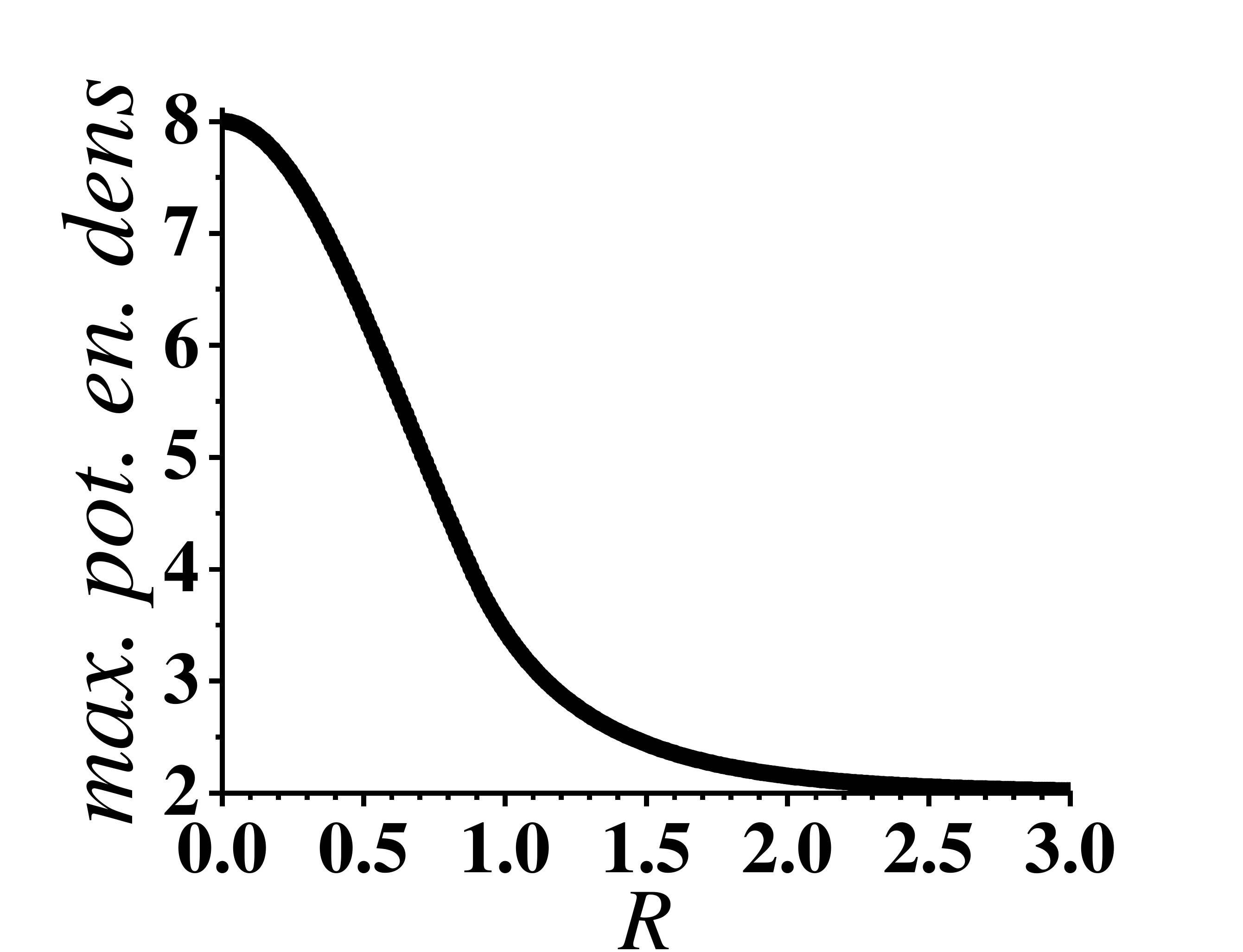}\label{fig:3kinksMaximumPotentialEnergyDensityV010}}
  \subfigure[\:gradient]{\includegraphics[width=0.24
 \textwidth]{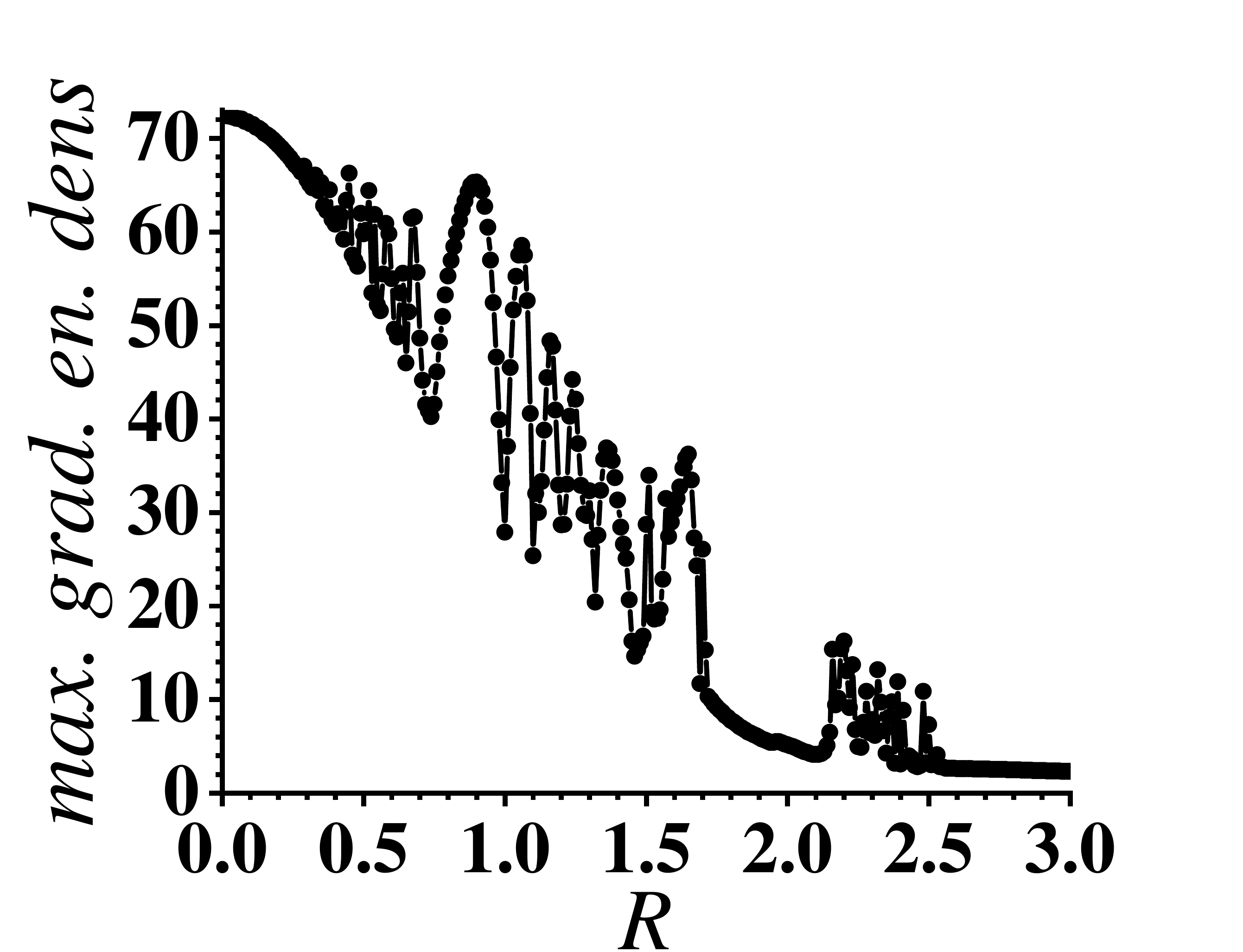}\label{fig:3kinksMaximumGradientEnergyDensityV010}}
 \\ 
\caption{Maximal energy densities as functions of $R$ for collisions of three kinks.} 
\label{fig:3kinksMaxEnergyDensites}
\end{figure}

We have obtained the dependencies of the maximal energy densities on the parameter $R$, see Fig.~\ref{fig:3kinksMaxEnergyDensites}.
\begin{figure*}[t!]
\begin{center}
  \centering
       \subfigure[\:$R=1.60$]{\includegraphics[width=0.3
 \textwidth]{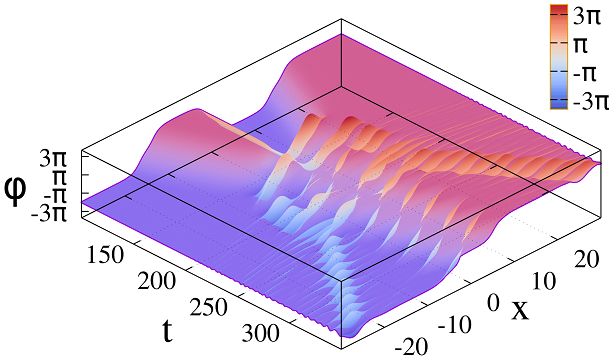}\label{fig:F3kR160V01X25}}
     \subfigure[\:$R=1.70$]{\includegraphics[width=0.3
 \textwidth]{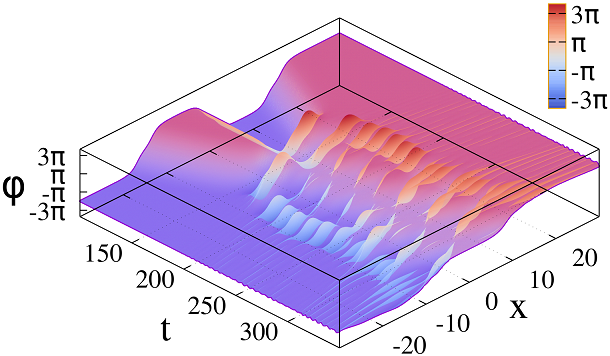}\label{fig:F3kR170V01X25}}
  \subfigure[\:$R=1.78$]{\includegraphics[width=0.3
 \textwidth]{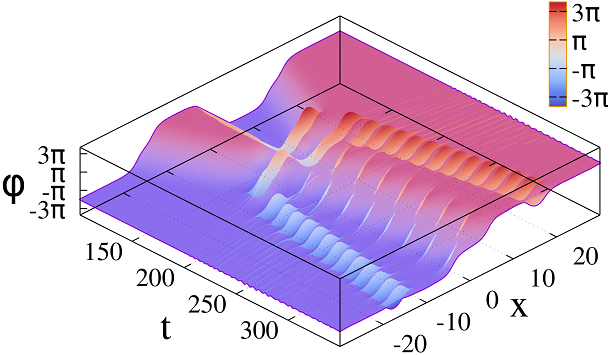}\label{fig:F3kR178V01X25}}
 \\
    \subfigure[\:$R=1.86$]{\includegraphics[width=0.3
 \textwidth]{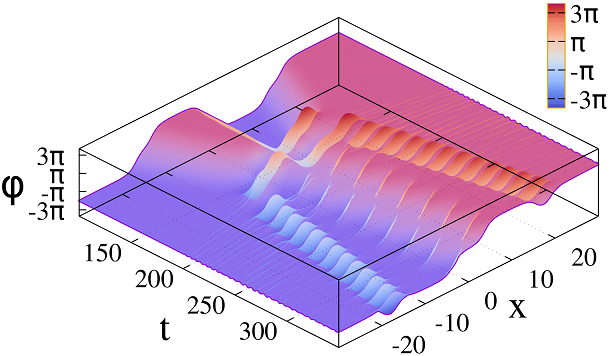}\label{fig:F3kR186V01X25}}
  \subfigure[\:$R=1.94$ ]
{\includegraphics[width=0.3
 \textwidth]{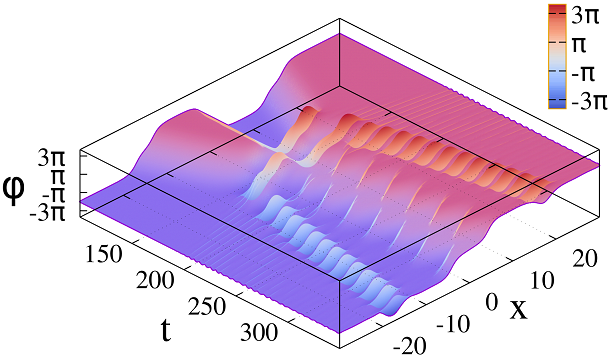}\label{fig:F3kR194V01X25}
}
  \subfigure[\:$R=2.02$]{\includegraphics[width=0.3
 \textwidth]{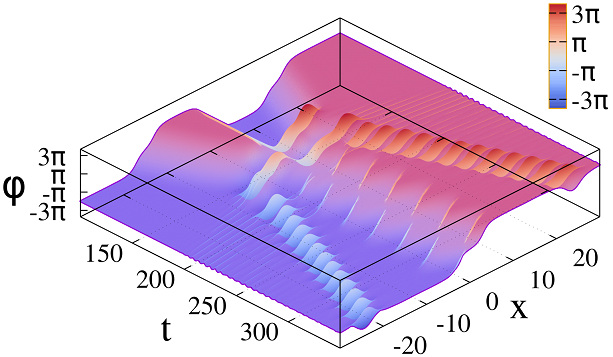}\label{fig:F3kR202V01X25}}
 \\
  \subfigure[\:$R=2.10$  ]
{\includegraphics[width=0.3
 \textwidth]{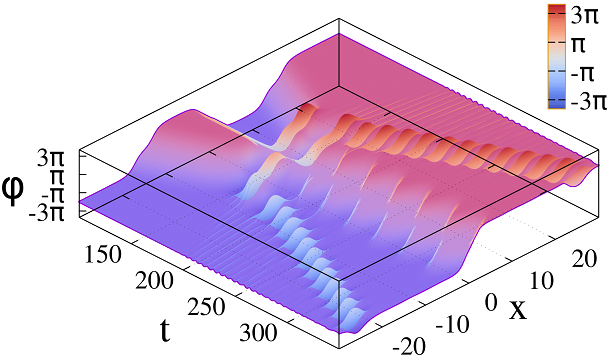}\label{fig:F3kR210V01X25}}
  \subfigure[\:$R=2.20$  ]
{\includegraphics[width=0.3
 \textwidth]{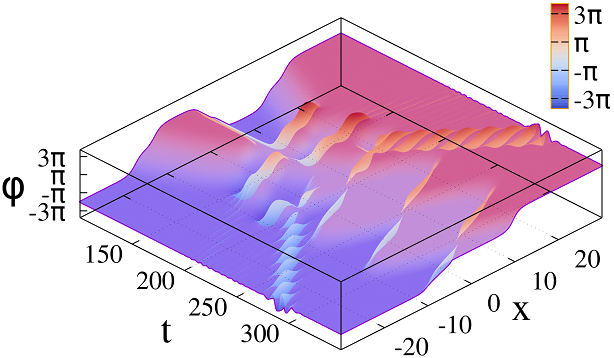}\label{fig:F3kR220V01X25}}
 \\
 \caption{Space-time picture of collisions of three kinks for some $R$'s from the range $1.6\le R\le 2.2$.}
  \label{fig:3kFields3}
\end{center}
\end{figure*}
The maximal potential energy density is monotonously decreasing, while the maximal kinetic, gradient and total energy densities behave in a complicated way, thereby reflecting complex interaction of the three colliding kinks.

Note that in the range $1.7\lsim R\lsim 2.1$ we see a smooth and monotonous segments of the maximal energy densities. We can assume that in this range of $R$ the field evolution is somewhat different. Actually, from Fig.~\ref{fig:3kFields3} we see that at these $R$'s two escaping oscillons in the final state are formed right after the kinks collision. At the same time, at $R\lsim 1.7$ and $R\gsim 2.1$ the escape of oscillons occur after some time during which the oscillons remain bound.


\subsection{Collision of two kinks and two antikinks}\label{sec:4kinks}

Finally, we studied the collision of four DSG kinks in one point. To do this, we used the initial configuration of the form of
$$
\phi_{\mathrm{k\bar k k\bar k}}(x) = \phi_{\mathrm{k}}\left(\frac{x+x_{02}^{}-v_\mathrm{in2}^{} t}{\sqrt{1-v_\mathrm{in2}^{2}}}\right) + \phi_{\mathrm{\bar k}}\left(\frac{x+x_{01}^{}-v_\mathrm{in1}^{} t}{\sqrt{1-v_\mathrm{in1}^{2}}}\right) +
$$
\begin{equation}\label{eq:DSG_4kinks}
+ \phi_{\mathrm{k}}\left( \frac{x-x_{01}^{}+v_\mathrm{in1}^{} t}{\sqrt{1-v_\mathrm{in1}^{2}}}\right) + \phi_{\mathrm{\bar k}}\left( \frac{x-x_{02}^{}+v_\mathrm{in2}^{} t}{\sqrt{1-v_\mathrm{in2}^{2}}}\right) - 2\pi,
\end{equation}
where $\phi_{\mathrm{k}(\mathrm{\bar k})}(x)$ is defined by Eq.~\eqref{eq:DSG_kinks_1} with $n=0$. The initial configuration \eqref{eq:DSG_4kinks} corresponds to two kinks and two antikinks moving to the collision point, see Fig.~\ref{fig:kakkak}. We used the following initial values of positions and velocities of the kinks and antikinks: $x_{01}=10$, $x_{02}=25$, $v_\mathrm{in1}^{}=0.05$, $v_\mathrm{in2}^{}=0.1$. This choice of the parameters guarantees almost simultaneous collision of all four waves in the origin $x=0$.

\begin{figure*}[t!]
\begin{center}
  \centering
     \subfigure[\:$R=0.50$]{\includegraphics[width=0.3
 \textwidth]{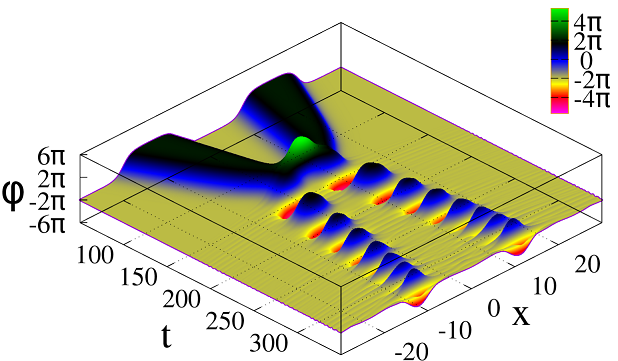}\label{fig:F4kR050V01V005X25X10}}
    \subfigure[\:$R=1.0$]{\includegraphics[width=0.3
 \textwidth]{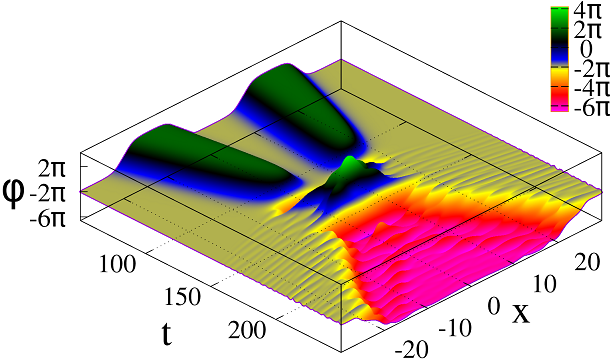}\label{fig:F4kR100V01V005X25X10}}
  \subfigure[\:$R=1.5$]{\includegraphics[width=0.3
 \textwidth]{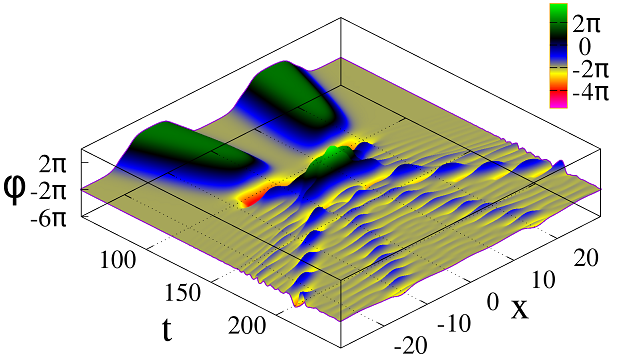}\label{fig:F4kR150V01V005X25X10}}
 \\
  \subfigure[\:$R=2.0$  ]
{\includegraphics[width=0.3
\textwidth]{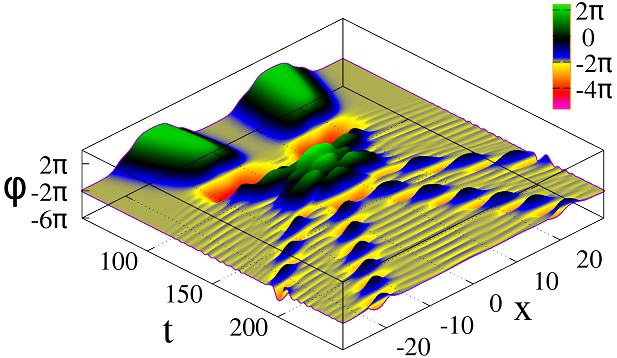}\label{fig:F4kR200V01V005X25X10}
}
  \subfigure[\:$R=2.5$]{\includegraphics[width=0.3
 \textwidth]{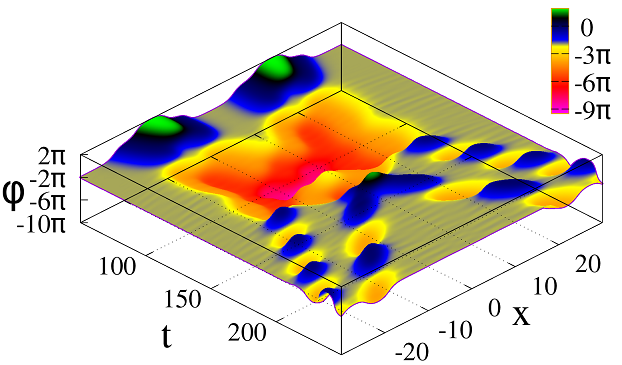}\label{fig:F4KR250V01V005X25X10}}
  \subfigure[\:$R=3.0$  ]
{\includegraphics[width=0.3
\textwidth]{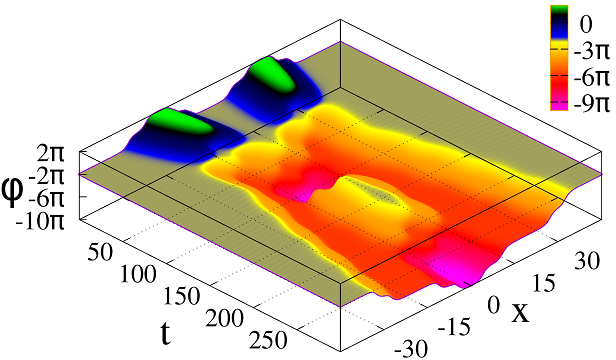}\label{fig:F4KR300V01V005X25X10}
}
 \\
 \caption{Space-time picture of collisions of four kinks for different $R$'s.}
  \label{fig:4kFields1}
\end{center}
\end{figure*}

The space-time picture of the collision for some selected values of the parameter $R$ is presented in Fig.~\ref{fig:4kFields1}.
We can see a vast variety of final states. At $R=1.5$, $R=2.0$ and $R=2.5$ we observe four oscillons in the final state. We could propose the following explanation of the formation of oscillons. Each DSG kink/antikink consists of two {\it subkinks}, see Eq.~\eqref{eq:DSG_kinks_2}. In the kinks collision the constituent subkinks interact with each other in a complicated way, thus forming bound states --- oscillons. This means that we consider an oscillon as a bound state of subkinks (at the same time, remind that a bion is a bound state of kink and antikink). At $R=0.5$ we observe formation of two bions. At $R=1.0$ the final state has the form of escaping kink and antikink together with two oscillons also escaping from the collision point. At $R=3.0$ in the final state we see a bion localized near the origin $x=0$ and escaping DSG kink and antikink. Maximal values of the energy densities in the case of four kinks collisions are highly dependent on the initial positions of the kinks. That is why we did not study the energy densities here but briefly summarized the observed final states.


\section{Discussion and conclusion}\label{sec:conclusion}

We have studied collisions of several kinks of the double sine-Gordon model in one point (in a small spatial region). We have performed numerical simulations of
\begin{itemize}
\item[(a)]
collisions of a kink and an antikink,
\item[(b)]
collisions of two kinks and an antikink in one point,
\item[(c)]
collisions of two kinks and two antikinks in one point.
\end{itemize}

{\it Collisions of two kinks} ({\it i.e.}\ kink-antikink scattering) are thoroughly investigated in literature \cite{Gani.EPJC.2018.dsg,Gani.PRE.1999,Campbell.PhysD.1986.dsg}. In this paper we focused on energy distributions in the kink-antikink collisions at various values of the parameter $R$. We have obtained $R$-dependences of the maximal (over $x$ and $t$) kinetic, gradient, potential and total energy densities, Fig.~\ref{fig:2kinksMaxEnergyDensites}. The maximal kinetic, gradient and total energy densities behave rather stochastic at $0.5\lsim R\lsim 2.5$, while at $R\lsim 0.5$ and $R\gsim 2.5$ smooth and monotonous segments are observed. We suppose that such behavior could be a consequence of the fact that the initial velocity of the kinks $v_\mathrm{in}^{}=0.1$ is less than the critical velocity $v_\mathrm{cr}^{}$, see \cite{Gani.EPJC.2018.dsg}. This, in turn, leads to kinks' capture and formation of a bion. Energy redistribution in a bion is complicated and can be viewed as chaotic. At the same time, at $R\lsim 0.5$ and $R\gsim 2.5$ we have $v_\mathrm{in}^{}>v_\mathrm{cr}^{}$, and colliding kinks bounce off each other after the first impact. In this case the complex energy redistribution does not take place, and $R$-dependences of the maximal energy densities are smooth and monotonous.

Besides, at $1.5\lsim R\lsim 1.6$ we also observed small smooth and monotonous segments of the maximal energy densities in Fig.~\ref{fig:2kinksMaxEnergyDensites}. In this range of the parameter, unlike other close values of $R$, in the final state we have two escaping oscillons formed after two oscillations of the kink-antikink bound state.

In {\it three kinks collisions} (in collisions of two kinks and an antikink) we observed various final states, in particular, similar to (i) excited kink ($R=0.5$, Fig.~\ref{fig:F3kR050V01X25}), (ii) excited kink and escaping oscillons ($R=1.0, 1.5, 2.0, 2.5$, Figs.~\ref{fig:F3kR100V01X25}--\ref{fig:F3kR250V01X25}), and (iii) two escaping kinks and an antikink at rest ($R=3.0$, Fig.~\ref{fig:F3KR300V01X25}). The maximal kinetic, gradient and total energy densities behave stochastic. At the same time, within the range $1.7\lsim R\lsim 2.1$ we observed smooth and monotonous segments of the curves. We found that it could be a consequence of the formation of two escaping oscillons in the final state right after the kinks collision. Notice that the maximal total energy density in three-kink collisions is considerably larger than in two-kink collisions, see Figs.~\ref{fig:2kinksMaxEnergyDensites} and \ref{fig:3kinksMaxEnergyDensites}. At the same time, the maximal potential energy density is nearly the same for the two- and three-kink collisions. It means that the collision process does not affect on the maximal potential energy density. It is interesting that similar situation was observed in the multi-kink collisions in the sine-Gordon model \cite{Moradi.EPJB.2017.multikink} (up to seven kinks collision) and in the collisions of odd number of kinks in the $\varphi^4$ (up to five kinks collision) \cite{Moradi.CNSNS.2017.multikink}, while in the collisions of even number of kinks in the $\varphi^4$ (up to four kinks collision) and in the $\varphi^6$ multi-kink collisions the maximal potential energy depends on the number of colliding kinks \cite{Moradi.JHEP.2017.multikink,Moradi.CNSNS.2017.multikink}.

Finally, we have studied {\it collisions of four kinks} (two kinks and two antikinks). The initial positions and velocities of the kinks were fit in such a way in order to provide almost simultaneous collision of all four waves in one point. After collision we observed a vast variety of the final states. Besides, the maximal energy densities strongly depend on the initial positions of the colliding kinks. That is why the analysis of the maximal energy densities requires detailed investigation of their dependences on parameters of the initial condition, which could be a subject of a separate publication. In this paper we have limited ourselves to the classification of the most typical final configurations. In particular, at $R=0.5$ in the final state we observed two oscillating structures which we classified as bound states of the DSG kink and antikink. At $R=1.0$ we observed escaping kink and antikink along with two oscillons --- bound states of subkinks, which are components of the DSG kink (antikink). At $R=1.5$ and $R=2.0$ --- only oscillons, at $R=3.0$ --- escaping antikink and kink, while the other kink-antikink pair forms a bion localized near the collision point.

So we see a vast variety of final states in collisions of more than two kinks. Moreover, the behavior of the maximal energy densities strongly depends on oscillating structures being formed in the final state.

In conclusion, we would like to mention the following issues which are beyond the scope of our paper, nevertheless, in our opinion, could become subject to future work.
\begin{itemize}
\item First, the multi-kink scattering within the field-theoretic models with polynomial potentials of 8th or higher degrees. As we have already mentioned in the Introduction, such models can have kinks with long-range interaction between them due to the kinks' power-law tails. This long-range interaction could lead to new collective phenomena in multi-kink systems.
\item Second, the multi-kink scattering within models with non-polynomial potentials, {\it e.g.}, sinh-deformed $\varphi^4$ or $\varphi^6$, could be of interest, and will be studied in a separate publication.
\end{itemize}


\section*{Acknowledgments}

The work of MEPhI group was supported by the MEPhI Academic Excellence Project (Contract No.\ 02.a03.21.0005, 27.08.2013). V.A.~Gani also acknowledges the support of the Russian Foundation for Basic Research under Grant No.\ 19-02-00971.



\vspace{10mm}

\hrule

\vspace{20mm}

\begin{figure}[h!]
\centering
\includegraphics[width=0.2\textwidth]{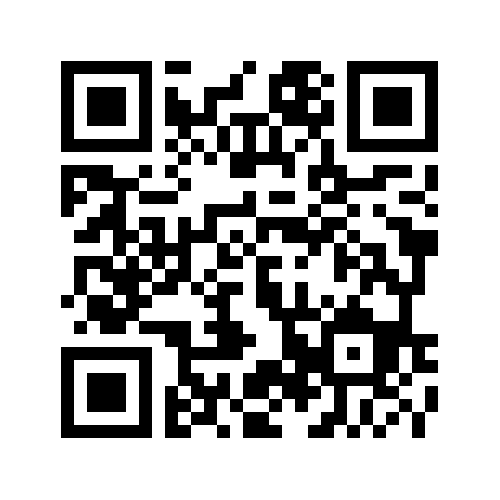}
\end{figure}

\end{document}